\documentclass[longauth]{aa} 

\usepackage{graphicx}
\usepackage{txfonts}
\usepackage{xcolor}
\usepackage{natbib}
\begin{document}

   \title{Looking into the faintEst WIth MUSE (LEWIS): on the nature of ultra-diffuse galaxies in the Hydra-I cluster}
   \subtitle{I. Project description and preliminary results}

   \author{Enrichetta Iodice\inst{1}\fnmsep\thanks{\email{enrichetta.iodice@inaf.it} }
          \and Michael Hilker  \inst{2}
          \and Goran Doll\inst{1,3}
          \and Marco Mirabile \inst{4,5}
          \and Chiara Buttitta \inst{1}
          \and Johanna Hartke \inst{6,7}
           \and Steffen Mieske \inst{8}
          \and Michele Cantiello\inst{4}
          \and Giuseppe D'Ago \inst{9}
          \and Duncan A. Forbes \inst{10}
          \and Marco Gullieuszik \inst{11}
          \and Marina Rejkuba \inst{2}
          \and Marilena Spavone \inst{1}
          \and Chiara Spiniello \inst{12}
          \and Magda Arnaboldi \inst{2}
          \and Enrico M. Corsini \inst{11,13}
          \and Laura Greggio \inst{11}
          \and Jesus Falc{\'o}n-Barroso \inst{14}
          \and Katja Fahrion \inst{15}
          \and Jacopo Fritz \inst{16}
          \and Antonio La Marca\inst{17,18}
          \and Maurizio Paolillo \inst{1,3}
          \and Maria Angela Raj \inst{17}
          \and Roberto Rampazzo \inst{11}
          \and Marc Sarzi \inst{19}
          \and Giulio Capasso \inst{1}
          }

           \institute{INAF $-$ Astronomical Observatory of Capodimonte, Salita Moiariello 16, I-80131, Naples, Italy
            \and
            European Southern Observatory, Karl$-$Schwarzschild-Strasse 2, 85748 Garching bei München, Germany
            \and
            University of Naples ``Federico II'', C.U. Monte Sant'Angelo, Via Cinthia, 80126, Naples, Italy
            \and
            INAF $-$ Astronomical Observatory of Abruzzo, Via Maggini, 64100, Teramo, Italy
            \and
            Gran Sasso Science Institute, viale Francesco Crispi 7, I-67100 L'Aquila, Italy
            \and
            Finnish Centre for Astronomy with ESO (FINCA), FI-20014 University of Turku, Finland
            \and
            Tuorla Observatory, Department of Physics and Astronomy, FI-20014 University of Turku, Finland            
            \and 
            European Southern Observatory, Alonso de Cordova 3107, Vitacura, Santiago, Chile
            \and
            Cambridge Astronomy Survey Unit (CASU), Institute of Astronomy, University of Cambridge, Cambridge, UK
            \and
            Centre for Astrophysics \& Supercomputing, Swinburne University of Technology, Hawthorn VIC 3122, Australia
        \and
            INAF $-$ Osservatorio Astronomico di Padova, Vicolo dell’Osservatorio 5, I-35122 Padova, Italy
        \and
            Department of Physics, University of Oxford, Denys Wilkinson Building, Keble Road, Oxford OX1 3RH, UK
            \and 
            Dipartimento di Fisica e Astronomia ``G. Galilei'', Universit\`a di Padova, vicolo dell'Osservatorio 3, I-35122 Padova, Italy
            \and 
        Instituto de Astrofísica de Canarias, Calle V\'ia L\'actea s/n, 38200 La Laguna, Spain
            \and 
        European Space Agency (ESA), European Space Exploration and Research Centre (ESTEC), Keplerlaan 1, 2201 AZ Noordwijk, The Netherlands
            \and 
        Instituto de Radioastronomía y Astrofísica, UNAM, Campus Morelia, A.P. 3-72, C.P. 58089, Mexico
            \and 
            Kapteyn Astronomical Institute, University of Groningen, Postbus 800, 9700 AV Groningen, The Netherlands
           \and
        SRON Netherlands Institute for Space Research, Landleven 12, 9747 AD Groningen, The Netherlands
           \and
            Armagh Observatory and Planetarium, College Hill, Armagh, BT61 9DG UK
             }

   \date{Received ....; accepted ...}

 
\abstract 
{Looking into the faintEst WIth MUSE (LEWIS) is an ESO large observing programme aimed at obtaining 
the first homogeneous integral-field spectroscopic survey of 30 extremely low-surface brightness (LSB) galaxies 
in the Hydra I cluster of galaxies, with MUSE at ESO-VLT. The majority of LSB galaxies in the sample 
(22 in total) are ultra-diffuse galaxies (UDGs). 
Data acquisition started in December 2021 and is expected to be concluded by March 2024.
{  Until June 2023, 29 targets were observed and the redshift derived for 20 of them.
The distribution of systemic velocities $V_{\rm sys}$ ranges between 2317~km/s and 5198~km/s
and is centred on the mean velocity of Hydra I ($V_{\rm sys} = 3683 \pm 46$~km/s). 
Considering the mean velocity and the velocity dispersion of the cluster ($\sigma_{\rm cluster} \sim 700$ km/s), 17 out of 20 targets are confirmed cluster members. The three objects with velocities more than $2\sigma_{\rm cluster}$ distant from the cluster mean velocity could be two background and one foreground galaxies.}
%
To assess the quality of the data and demonstrate the feasibility of the science goals, 
we report the preliminary results obtained for one of the sample galaxies, UDG11.
For this target, we {\it i)} derived the stellar kinematics, including the 2-dimensional maps of 
line-of-sight velocity and velocity dispersion, {\it ii)} constrained age and metallicity, 
and {\it iii)} studied the globular cluster (GC) population hosted by the UDG.
Results are compared with the available measurements for UDGs 
and dwarf galaxies in literature. 
By fitting the stacked spectrum inside one effective radius, we find that UDG11 
has a velocity dispersion $\sigma = 20 \pm 8$~km/s, 
it is old ($10\pm1$~Gyr), metal-poor ([$M$/H]$=-1.17\pm0.11$~dex) and has a {  total dynamical 
mass-to-light ratio M$/L_V\sim 14$}, comparable to those observed for classical dwarf galaxies. 
The spatially resolved stellar kinematics maps suggest that UDG11 does not show a 
significant velocity gradient along either major or minor photometric axes, and 
the average value of the velocity dispersion is $\langle \sigma \rangle_{\rm e} = 27 \pm 8$~km/s.
We find two GCs kinematically associated with UDG11. The estimated total number of GCs in UDG11, corrected for the spectroscopic completeness limit, is $N_{\rm GC}= 5.9^{+2.2}_
{-1.8}$, which corresponds to a GC specific frequency of $S_N = 8.4^{+3.2}_{-2.7}$.
%
}


\titlerunning{The LEWIS project}
\authorrunning{E. Iodice et al. }
\maketitle

%

\section{Introduction}\label{sec:intro}

Ultra-diffuse galaxies (UDGs) are among the faintest and lowest surface brightness 
($\mu_{0,g} \geq 24$ mag arcsec$^{-2}$, $R_{\rm e}\geq 1.5$~kpc) galaxies known in the universe. 
Very low surface brightness (LSB) galaxies are known since the 80s, they were discovered 
in the Virgo and Fornax clusters decades ago \citep{Sandage1984,Impey1988,Ferguson1988,Bothun1991}.
The term UDG was introduced by \citet{vanDokkum2015}, who detected several galaxies, including an extremely faint and diffuse galaxy, named DF44, in the Coma cluster, having an effective radius $R_{\rm e}\sim 4.3$~kpc, similar to that of the Milky-Way ($R_{\rm e}\sim4.5$~kpc), but with a stellar mass 100 times smaller. 

To survive cluster tides, being so diffuse and faint, UDGs should host a large amount of dark matter \citep{vanDokkum2015}.
This idea triggered an ever-increasing attention to the detection and study of UDGs and gave to
these galaxies a special role in the realm of the LSB universe. Given the extremely 
low baryonic mass density, UDGs are indeed considered particularly suitable laboratories to test the formation 
of galaxies in the $\Lambda$-Cold Dark Matter ($\Lambda$CDM) framework.

Since 2015, several observational campaigns have been carried out to get deep images, mapping different 
environments from groups to clusters of galaxies, which provided large samples of LSB galaxies, 
including UDGs \citep{Yagi2016,vanderBurg2017,Trujillo2017,Venhola2017,Janssens2019,Mancera-pina2019udg,Prole2019b,Roman2019,Lim2020,Marleau2021,LaMarca2022b,Zaritsky2022}.

To classify an LSB galaxy as UDG, the most conservative approach is based on 
the empirical definition proposed by \citet{vanDokkum2015}. 
This requires UDGs to have a central surface brightness fainter than 24 mag arcsec$^{-2}$ (in the $g$ band) and
an effective radius larger than 1.5~kpc. However, other criteria have also been proposed which used 
different cuts in size and/or surface brightness limits \citep{Koda2015,Yagi2016,vanderBurg2017,ManceraPina2019}. 
In particular, taking advantage of a large and statistically significant sample of dwarf 
and LSB galaxies in the Virgo cluster, \citet{Lim2020} found that UDGs can be classified as the extremes in the broad scaling 
relationships between photometric and structural properties of LSB galaxies (e.g., total luminosity vs. $R_{\rm e}$ and $\mu_{\rm e}$).
UDGs turn out to be $\sim2.5\sigma$ fainter and larger than the average distribution of the parent dwarf galaxy sample.
These results support the idea that UDGs can be considered as extreme LSB tail of the size-luminosity distribution of 
dwarf galaxies.

The considerable amount of imaging data collected to date for UDGs has shown that these galaxies span a 
wide range of structural and photometric properties. 
Based on their integrated colours, it seems that two populations of UDGs exist. Red UDGs are mainly found 
in clusters of galaxies, while bluer objects are discovered in the low-density regions, 
i.e. the outskirts of clusters and in the field
\citep[see e.g.][]{Roman2017a,Leisman2017,Prole2019a,Marleau2021}. Red UDGs are also found 
in groups of galaxies \citep{Marleau2021}.

Deep images have also allowed the detection and study of globular cluster (GC) populations in UDGs.
By deriving the GC specific frequency \citep[$S_N = N_{\rm GC} \times 10^{0.4(M_V+15)}$,][]{Harris1981Sn}, 
observations from space and ground-based telescopes have revealed an extreme degree of variability of $S_{N}$ values.
Some UDGs are consistent with having no GCs population, while others present $S_{N}$  as high as $\sim$150 \citep{Prole2019b,Saifollahi2021,Saifollahi2022,Marleau2021,LaMarca2022b}.
The unusually high $S_N$ in some UDGs triggered a debate about the fraction of DM in UDGs.
Assuming that the relation between the total number of GCs, $N_{\rm GC}$, and the host galaxy's halo virial mass, 
valid from giant to dwarf galaxies \citep[see][and references therein]{Burkert2020}, also holds in the LSB regime,
UDGs with large $N_{\rm GC}$ values might be DM-dominated systems, with halo masses 100 times more massive 
($M_{\rm h} \geq 10^{11}$~ M$_{\odot}$) than higher surface brightness dwarf galaxies of similar luminosity. 

To date, the DM content of the UDGs is a highly debated topic. 
The few spectroscopic studies, focusing on special cases, point to a rather diverse population.
Some UDGs were found to host a very massive DM halo \citep{Toloba2018,vanDokkum2019,Forbes2021,Gannon2021}, 
while at the same time, there are some UDGs with a ``normal'' DM halo, i.e. having the DM content 
consistent with that of other dwarf galaxies of similar luminosity. 
Finally, a few UDGs have also been discussed to populate the opposite extreme, being basically DM-free
\citep{vanDokkum2018,Collins2021}.

Because of their LSB nature, getting spectroscopic data for UDGs is a challenging task. 
To date,
as opposed to the deep images available, we still lack a statistically significant sample of UDGs with spectroscopy, 
which strongly limits our constraints and conclusions on their stellar populations and DM content. 
For only two to three dozen of UDGs, the available spectroscopic studies 
reveal the existence of both, metal-poor ($-0.5 \leq [M$/H$] \leq -1.5$ dex) and old systems 
\citep[$\sim$9 Gyr; e.g.,][]{Ferre-Mateu2018,Pandya2018,Fensch2019}, 
as well as younger star-forming UDGs \citep{Martin-Navarro2019}. 
The kinematic measurements of UDGs available in groups and clusters 
suggest that the rotation velocity of stars is very low
\citep[see][and references therein]{Gannon2023}.

The wide range of photometric and spectroscopic properties, including the GCs content, do not fit into a single 
formation scenario, and there is a general consensus that different formation channels can be invoked to form 
galaxies with UDG-like properties. 
\citet{vanDokkum2015} coined the term ``failed'' galaxies, suggesting that these objects
with high DM content and large effective radii might have lost gas supply at an early epoch,
evolving to become quenched and diffuse galaxies.
Nowadays, a plethora of UDG formation mechanisms have been proposed, which 
are nicely able to form a galaxy with the typical morphology of UDGs, but they predict
different DM amounts, age and metallicity, and gas content. 
It has to be shown which formation scenarios are most consistent with the 
observational properties of UDGs.

For simplicity, formation scenarios for UDGs can be divided into two groups, based on the physical processes at work: 
internal and external mechanisms. 

Star-formation feedback and highly rotating DM halos are both possible internal mechanisms that can form large and 
diffuse galaxies. In the former case, repeated star formation episodes during early galaxy evolution can drive the gas 
out to large radii and prevent the subsequent star formation \citep{diCintio2017}. 
In the latter case, the high specific angular momentum of a DM halo prevents gas from effectively collapsing into a dense structure
\citep{Amorisco2016,Rong2017,Tremmel2019}.
In both scenarios, the resulting UDGs are gas-rich and have a ``normal'', dwarf-like DM halo.

\begin{figure*}
    \centering
    \includegraphics[width=\hsize]{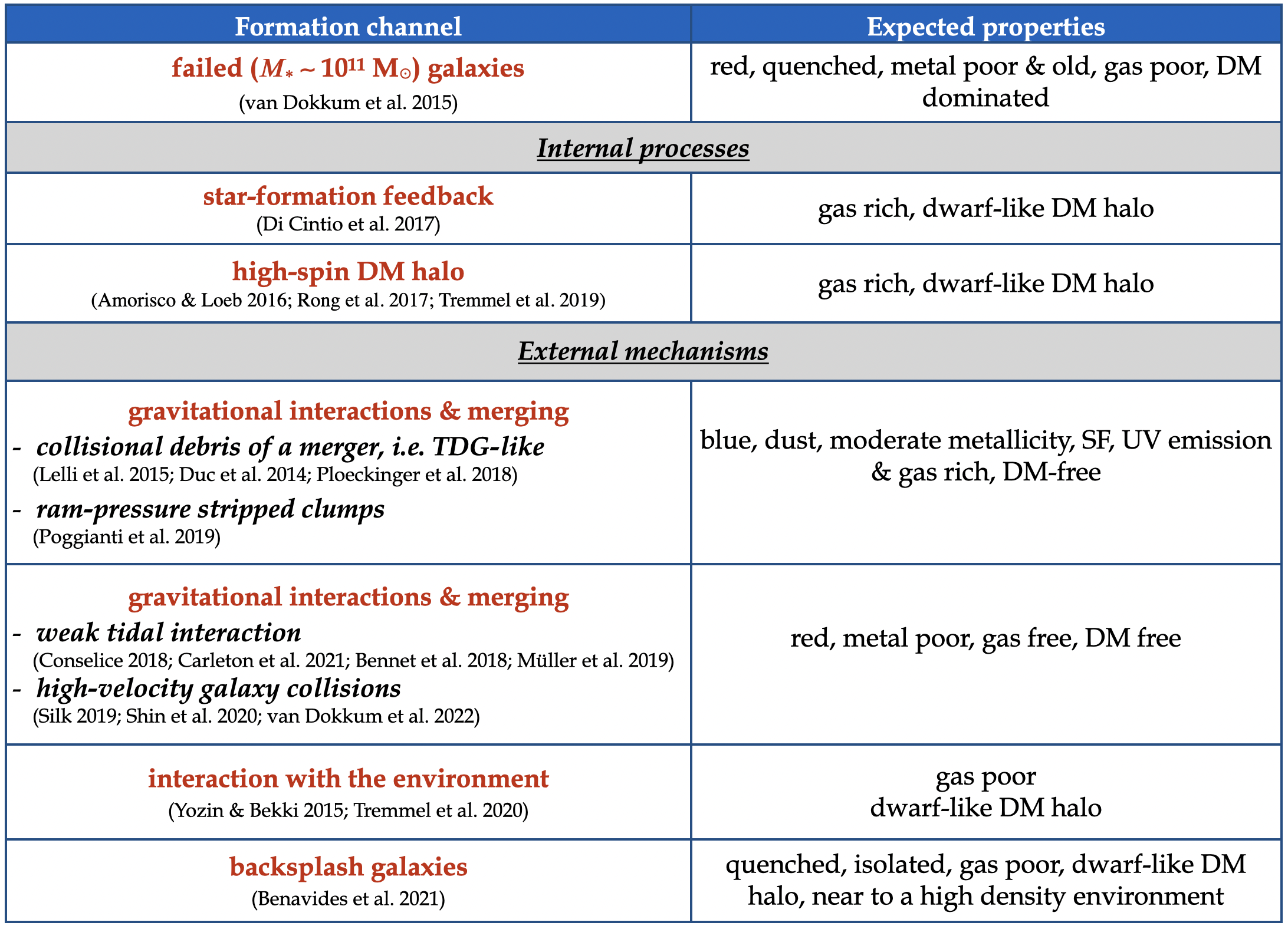}
    \caption{Schematic view of the main formation channels proposed for UDGs. In the left boxes are listed the relevant physical processes. The predicted properties for UDGs, from each formation channel, are reported in the boxes on the right.}
    \label{fig:formation}
\end{figure*}

Gravitational interactions and merging between galaxies, as well as interactions with the environment, are external processes that
might shape galaxies to become UDG-like.
Similar to the tidal dwarf galaxies (TDGs), UDGs might originate from the collisional debris of a merger 
\citep{Lelli2015,Duc2014,Ploeckinger2018}. \citet{Poggianti2019} suggested that UDGs might form from 
ram pressure-stripped gas clumps in the extended tails of infalling cluster galaxies. 
Both scenarios predict
blue, dusty, star-forming, and DM-free UDGs, with moderate to low metallicity and UV emission.
Weak tidal interaction of a dwarf galaxy with a massive nearby giant galaxy has also been addressed as a possible
mechanism to form a UDG \citep{Conselice2018,Carleton2021,Bennet2018,Muller2019,Gannon2021}. 
High-velocity galaxy collisions might generate several debris, where some of them could remain gravitationally-bound systems with a UDG-like structure
\citep{Silk2019,Shin2020,vanDokkum2022}. Also in these latter cases, the formed UDGs are expected to be 
DM-free galaxies, but red and gas-poor systems.
UDGs could also form from large dwarf galaxies, which, during their interaction with the cluster environment, 
got their gas removed by ram pressure stripping, and thus
halting subsequent star formation \citep{Yozin2015,Tremmel2020}. 
The resulting UDG is gas-poor and has a dwarf-like DM content.
Finally, in the framework of external processes, 
a quenched, isolated, gas-poor UDG, with a dwarf-like DM halo, might result as a backsplash galaxy. 
In this case, former satellites of a group or cluster halo in an early epoch, are now found a 
few megaparsecs  away from them \citep{Benavides2021}.
All the above scenarios and related predicted properties for UDGs are summarised in Fig.~\ref{fig:formation}.

Based on the IllustrisTNG simulations, \citet{Sales2020} proposed two different formation channels for cluster UDGs. 
A population of ``Born-UDGs'' (B-UDGs) could form in the field and later enter the cluster environment. The B-UDGs 
originate from LSB galaxies that, once having joined the cluster potential, lost their gas supply and quench. Differently,
tidal forces could act on luminous galaxies in the cluster, removing their DM and puffing up their stellar component.
As a consequence, these galaxies evolve into UDGs, and they are named ``Tidal-UDGs'' (T-UDGs). 
T-UDGs populate the centre of the clusters and, at a given stellar mass, have lower velocity dispersion, higher metallicity and lower DM fraction with respect to the B-UDGs.

In summary, observations strongly suggest that the class of UDGs might comprise different types of galaxies, with 
different intrinsic properties (e.g., colours, stellar populations, and DM fractions). 
Theoretical works on UDGs, also reviewed above, show that more than one formation channel might exist to account
for the different types of UDGs, or, reasonably, a combination  of physical processes to account also for environmental effects.
The lack of stellar kinematics and stellar population properties is the main limitation to provide stringent 
conclusions on the nature of UDGs and to discriminate between the formation channels.

In this paper, we present the ``Looking into the faintEst WIth MUSE'' (LEWIS) project, which aims at obtaining the first homogeneous integral-field
spectroscopic survey of 30 extreme LSB galaxies, including UDGs, in the Hydra\,I cluster of galaxies, with MUSE at ESO-VLT. 
Doubling the number of spectroscopically studied UDGs, with this project we will make a decisive impact in this field. 
With LEWIS we will map, for the first time, the stellar population and DM content of a complete sample of UDGs in a 
galaxy cluster, based on spectroscopic data.

The paper is organised as follows. The galaxy sample and science goals of the LEWIS project are presented in 
Sec.~\ref{sec:lewis}. Observations and data reduction are described in Sec.~\ref{sec:data}. The redshift estimates 
for all the so far observed sample UDGs are provided in Sec.~\ref{sec:redshift}.
In Sec.~\ref{sec:data_quality} we show the analysis of the MUSE data, with a detailed description for 
one of the galaxies in the sample, UDG11, chosen as test case.
The preliminary results are discussed in Sec.~\ref{sec:disc}, and conclusions are provided in Sec.~\ref{sec:concl}.

\section{Galaxy sample and science goals of the LEWIS project}\label{sec:lewis}

\begin{figure*}
    \centering
    \includegraphics[width=\hsize]{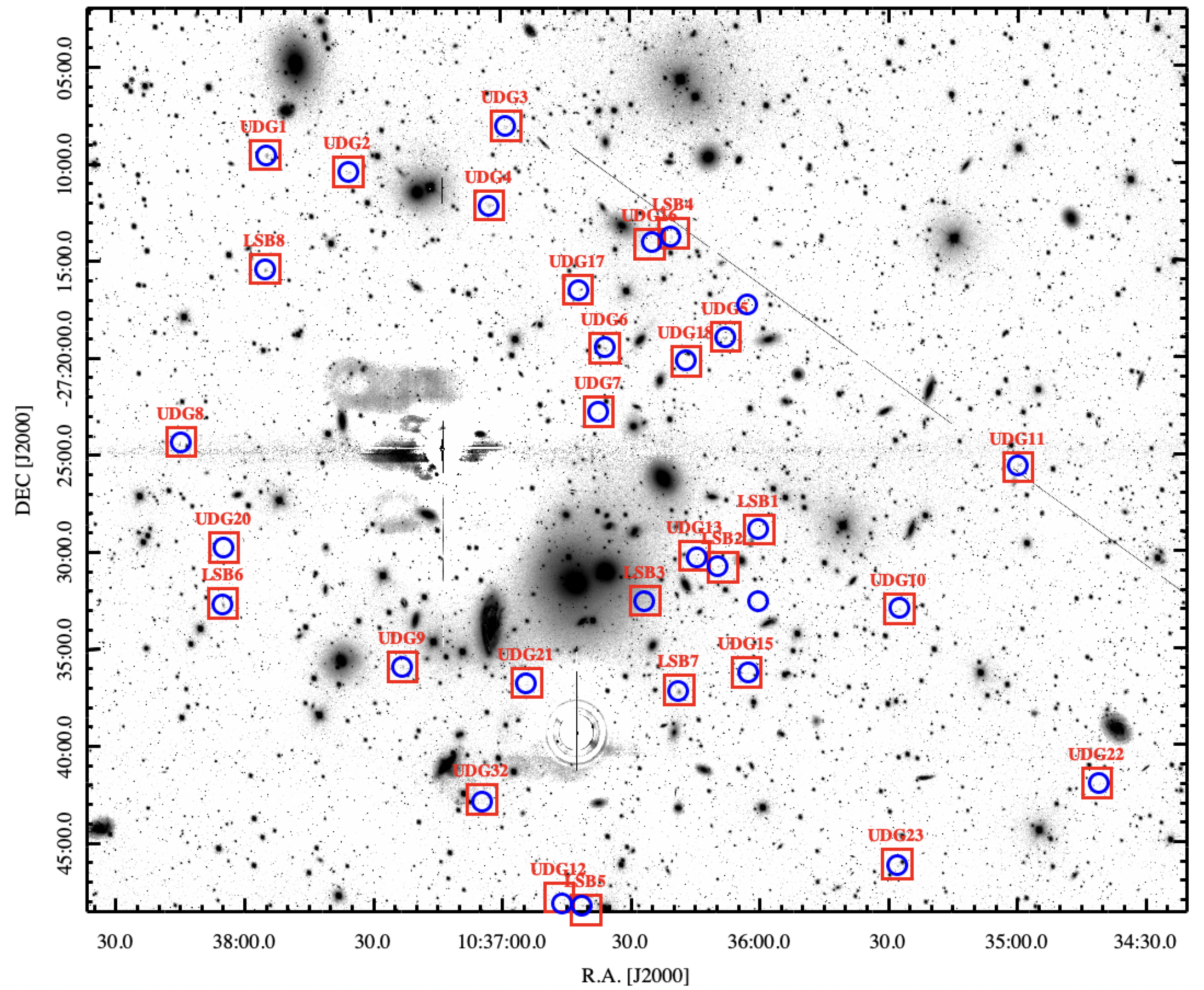}
    \caption{Optical $g$-band mosaic from VST of the \object{Hydra\,I} cluster
    ($56.7\arcmin \times 46.55$~\arcmin $\sim0.8 \times 0.7$~Mpc). North is up and East to the left. The two bright stars, close to the cluster core, where modelled and subtracted from the image, as explained in \citet{Iodice2020b}.
    The 32 new UDGs and LSB galaxies detected by \citet{Iodice2020b} and \citet{LaMarca2022b} 
    are marked as blue circles. Red boxes show the LEWIS sample presented in this work and listed in Table~\ref{tab:UDGsample}.}
    \label{fig:Hydra}
\end{figure*}

LEWIS is an ESO Large Programme, started in 2021, approved during the ESO period 108 (P.I. E. Iodice, ESO programme ID 108.222P), 
to obtain the first homogeneous integral-field spectroscopic survey of UDGs in the Hydra\,I cluster of 
galaxies (see Fig.~\ref{fig:Hydra}).
This is a rich environment of galaxies, located in the southern hemisphere at a distance of $51\pm6$ Mpc 
\citep{Christlein2003}, with a virial mass of $2 \times 10^{14}$~M$_{\odot}$ \citep{Girardi1998}, a virial radius $R_{\rm vir}\sim 1.6$~Mpc, and a velocity dispersion of $\sigma_{\rm cluster} \simeq700$~km/s \citep{Lima-dias2021}.
Hydra\,I has been extensively studied using deep images and multi-object spectroscopy \citep[e.g.,][]{Misgeld2008,Misgeld2011,Richtler2011,Arnaboldi2012,Hilker2018, Barbosa2018,Barbosa2021b}.
Main results show that this cluster is still in an active phase of mass assembly, since ongoing interactions are detected 
around the brightest cluster member \object{NGC\,3311} \citep[see][and references therein]{Barbosa2018,Iodice2021}.
The projected distribution of all cluster members, bright galaxies and dwarfs, shows three main over-densities:
the core of the cluster, around NGC~3311 and NGC~3309, a sub-group of galaxies elongated North-South and a sub-group 
of galaxies in the South-East \citep{LaMarca2022b}.

The latest Hydra I catalogue presented 317 galaxies fainter than $M_r>-18.5$~mag and semi-major axis 
larger than ~200 pc, of which about 230 new candidates have been recently discovered by
\citet{Iodice2020b,Iodice2021} and \citet{LaMarca2022a,LaMarca2022b}. 
The authors studied the photometric properties of this class of objects, 
which are briefly summarised below.
According to the colour-magnitude relation for early-type giant and dwarf 
galaxies in Hydra\,I \citep{Misgeld2008}, all of the new candidates are consistent with being cluster members.
%
In this sample, according to the definition proposed by \citet{vanDokkum2015}, i.e. $R_{\rm e} \geq 1.5$~kpc and $\mu_{0,g} \geq 24$~mag/arcsec$^2$, 22 objects are classified as UDGs \citep{Iodice2020b,Iodice2021,LaMarca2022b}. 
Additional 10 galaxies are very extended ($R_{\rm e} \geq 1.5$~kpc), but with 
$\mu_{0,g}\geq 23$~mag/arcsec$^2$, were classified as LSB dwarfs.
{  Taking into account the virial mass of $\sim10^{14}$\,M$_{\odot}$ for the Hydra I cluster \citep[see Fig. 6 in][]{LaMarca2022b}, and the UDG abundance-halo mass relation \citep[$N_{\rm UDG}\propto M_{200}^{1.11}$,][]{vanderBurg2017}, the expected number of  UDGs in Hydra I within  $1R_{\rm vir}$ is $48\pm10$~UDGs. Therefore, the detection of 22 UDGs inside $\simeq0.4R_{\rm vir}$ of the Hydra~I cluster can be considered a complete sample for this class of objects.}
%
Based on photometric and size selection, GC candidates are identified around a few of those LSB 
galaxies, with a total number of GCs per galaxy $N_{\rm GC} \geq 2$.

The newly discovered LSB dwarfs and UDGs span a wide range of central surface brightness 
($23 \leq \mu_{0,g} \leq 27$~mag/arcsec${^2}$) and effective radius ($1\leq R_{\rm e} \leq 4$~kpc). 
Compared to the population of early-type dwarf galaxies in the cluster, they have
similar integrated $g-r$ colours, $0.4 \leq g-r \leq 0.9$~mag, and stellar masses $M_{\star} = 5\times10^6 - 2\times10^8$~M$_{\odot}$.
Inside $\simeq 0.4 R_{\rm vir}$ of the Hydra\,I cluster, the structural and photometric parameters (i.e., surface brightness, size, color, and Sersic n-index) and GC content of all LSB galaxies have similar properties and trends to those observed for dwarf galaxies. 
Therefore, as addressed by \citet{LaMarca2022b}, these findings
suggest that a single population of LSB galaxies is present in this region of the cluster, and UDGs can be reasonably considered
as the extreme LSB tail of the size–luminosity distribution of all dwarfs in this environment. 
Finally, the LSB galaxies share a similar 2D projected distribution as observed for the dwarf and giant galaxies in the cluster: 
over-densities are found in the cluster core and north of the cluster centre. 
Similar results are found for other galaxy clusters,
where over-densities of UDGs are observed close to subgroups of other cluster members \citep[see e.g.][]{Janssens2019}. 
Observing UDGs spatially associated with groups infalling onto the cluster, would further support the idea that they might follow two formation paths, as proposed by \citet{Sales2020}.
As a conclusive remark, the previous results and ongoing studies on UDGs in Hydra\,I suggest 
that this environment offers a unique 
opportunity to analyse this class of LSB galaxies in great detail. 

The UDGs' nature and formation can be addressed by measuring their kinematics, stellar population and DM content 
as a function of their location in the cluster. Compared to the Coma cluster, the so far best-studied 
environment where UDGs have been investigated, Hydra~I is at half of Coma's distance and 
it is 10 times less massive. Therefore, Hydra~I offers the exquisite opportunity to analyse LSB galaxies,
including UDGs, in an environment of
different mass scales, and relate their properties to the mass assembly processes. 
In particular, if the new spectroscopic data from the LEWIS project confirm 
the asymmetric distribution of UDGs, we can investigate whether the galaxies in the subgroups have different properties 
from those in the outskirts of the cluster, indicating that the latter systems have formed as genuine UDGs different 
from those in the denser inner environment.

Given the large variety of observed properties (mainly based on deep images) and theoretical predictions, 
 the LEWIS project will provide a notable boost 
in our knowledge of UDG structure and formation in a cluster environment \citep[see][]{Forbes2023}.
In particular, we expect to address the following science goals, which are the main debated issues on the nature of UDGs:
\begin{itemize}
    \item DM content in each UDG of the sample through dynamical mass estimates from stellar kinematics. 
    Since the UDGs are not uniformly distributed inside the cluster, we will check if the DM content correlates with 
    the environment in which the UDG resides;
    \item UDGs’ star formation history from SED fitting of their integrated spectra. This allows to study the evolutionary link between the UDGs and other dwarf galaxies through a comparison of their stellar population 
    and structural properties;
    \item The spectroscopic confirmation of GC candidates around UDGs will improve their ${S_N}$ estimates, which in turn will put on a firmer basis the discussion about possible overdensities of GCs around some UDGs and the relation to the host galaxy DM content.
\end{itemize}

To these aims, the main objectives of LEWIS are to derive the stellar kinematics, stellar populations and the 
spectroscopic specific frequency ${S_N}$ of the hosted GCs, for all the selected galaxies in our sample. 
As stated in Sec.~\ref{sec:intro}, similar studies are available only for about 35 UDGs in total, mainly in the Coma cluster \citep{vanDokkum2017,Ruiz-Lara2018,Ferre-Mateu2018,Gu2018}. 
In particular, integral-field (IF) spectroscopy is available only for about 
a dozen UDGs \citep{Martin-Navarro2019,Emsellem2019,Muller2020,Gannon2021,Webb2022,Gannon2023}.

From the sample of 32 LSB galaxies, photometrically detected in the Hydra~I cluster \citep{Iodice2020b,LaMarca2022b}, we have selected 30 objects (22 UDGs and 8 LSB galaxies) for the spectroscopic follow-up with MUSE at ESO-VLT within our LEWIS project. They were selected to have an effective surface brightness in the range $25 \leq \mu_{\rm e} \leq 27.5$~mag/arcsec$^2$ in the $g$ band, which provides the minimum signal-to-noise ratio (SNR$\sim$5-10, depending on the surface brightness) per spaxel
in a reasonable integration time ($\sim$2-6 hours), required for the main goals of this project.
{  The two targets excluded from the spectroscopic LEWIS follow-up, UDG14 and UDG19 in \citet{LaMarca2022b}, are the faintest objects of the photometric sample with $\mu_{\rm e}=28.5$~mag/arcsec$^2$ in the $g$ band. According to the
color-magnitude relation \citep[see Fig. 3 in][]{LaMarca2022b}, both galaxies can be considered as Hydra I cluster members.}
Galaxies in the LEWIS sample are listed in Table~\ref{tab:UDGsample} and shown in Fig.~\ref{fig:Hydra}. 





\begin{table*}
\renewcommand{\arraystretch}{1.}
\setlength{\tabcolsep}{3pt}
\small 
\caption{LEWIS sample: UDG and LSB galaxies in the Hydra I cluster.} 
\label{tab:UDGsample}
\vspace{15pt}
\begin{tabular}{lccccccccccc}
\hline\hline
Object & R.A. & DEC & $M_r$ & $M_*$ & $\mu_{\rm e}$ & $\mu_{0}$ &  $R_{\rm e}$ & $N_{\rm GC}$ & Obs. Status & Exp. Time & $V_{\rm sys}$\\ 
 &[J2000] & [J2000] & [mag] & [$10^{8}$~M$_\odot]$ & [mag/arcsec$^{2}$] &[mag/arcsec$^{2}$] & [kpc]& & & [hrs] & [km/s] \\
 (1) & (2) & (3) & (4) & (5) & (6) & (7) & (8) & (9) & (10) & (11) & (12) \\
\hline \vspace{-7pt}\\
UDG 1 &10:37:54.12 & -27:09:37.50& -15.48 &  1.12 & 25.2$\pm$0.1 & 24.2$\pm$0.1 & 1.75$\pm$0.12 & 0$\pm$1 & C & 2.00 & 4219$\pm$9 \\ 
UDG 2 & 10:37:34.89 & -27:10:29.94 & -14.27 &  0.55 & 26.2$\pm$0.1 & 25.0$\pm$0.1 & 1.55$\pm$0.12 & 7$\pm$3 & P & 1.33 & -\\
UDG 3 & 10:36:58.63 & -27:08:10.21 & -14.70 &  1.65 & 26.1$\pm$0.2 & 25.2$\pm$0.2 & 1.88$\pm$0.12 & 15$\pm$6 & C & 3.00 & 3550$\pm$26\\
UDG 4 & 10:37:02.64 & -27:12:15.01 & -16.03 & 10.6 & 25.8$\pm$0.1 &24.9$\pm$0.1 & 2.64$\pm$0.12 & 2$\pm$1 & C & 2.00 & 2317$\pm$13 \\
UDG 5 & 10:36:07.68 & -27:19:03.26 & -14.66 &  1.16 & 25.3$\pm$0.3 &23.7$\pm$0.3 & 1.42$\pm$0.12 & 0$\pm$1 & S & - & -\\
UDG 6 & 10:36:35.80 & -27:19:36.12 & -14.38 &  0.32 & 25.3$\pm$0.1 &24.1$\pm$0.1 & 1.37$\pm$0.12 & 0$\pm$1 & P & 0.75 & -\\
UDG 7 & 10:36:37.16 & -27:22:54.93& -13.72  &  0.49 & 26.9$\pm$0.4 &24.4$\pm$0.4 & 1.66$\pm$0.12 & 3$\pm$1 & C & 3.80 & 4134$\pm$28\\
UDG 8 & 10:38:14.59 & -27:24:27.07 & -14.87 &  0.53 & 25.0$\pm$0.6 &23.2$\pm$0.6 & 1.40$\pm$0.12 & 0$\pm$1 & P & 0.75 & {  4793$\pm$16}\\
UDG 9 & 10:37:22.85 & -27:36:02.80 & -15.16 &  1.78 & 26.8$\pm$0.2 &24.2$\pm$0.2 & 3.46$\pm$0.12 & 7$\pm$1 & C & 3.90 & 4302$\pm$13\\
UDG 10 & 10:35:27.32 & -27:33:03.86 & -13.89 &  0.26 & 27.3$\pm$0.3 &24.3$\pm$0.3 & 2.29$\pm$0.10 & 0$\pm$1 & C & 3.85 & 3577$\pm$139\\
UDG 11 & 10:34:59.55 & -27:25:37.95 & -14.75 &  0.63 & 25.7$\pm$0.1 &24.4$\pm$0.1 & 1.66$\pm$0.12 & 7$\pm$3 & C & 6.10 & 3507$\pm$3\\
UDG 12 & 10:36:45.55 & -27:48:12.73 & -14.30 &  1.19 & 26.2$\pm$0.2 &25.1$\pm$0.2 & 1.64$\pm$0.12 & 0$\pm$1 & P & 1.50 & 4615$\pm$13\\
UDG 13 &10:36:14.49& -27:30:26.60&	 -12.73  & 0.20 & 27.3$\pm$0.1 & 24.2$\pm$0.2 & 1.60$\pm$0.20 & -  & C & 3.93 & 3438$\pm$52\\
UDG 15 & 10:36:02.55& -27:36:19.57 &  -11.95& 0.06 & 27.7$\pm$0.2 & 25.0$\pm$0.3 & 1.51$\pm$0.15 &  2$\pm$3 & C & 4.68 & 3553$\pm$88 \\
UDG 16 & 10:36:25.30 & -27:14:14.74 & -12.84& 0.11 & 27.6$\pm$0.1& 25.9$\pm$0.2 & 1.75$\pm$0.12&  - & S & - & -\\ 
UDG 17 & 10:36:41.72 & -27:16:37.48 &  -13.99& 1.20 & 26.7$\pm$0.1 & 24.9$\pm$0.1 & 1.50$\pm$0.20 & 3$\pm$3  & P & 1.50 & -\\
UDG 18 & 10:36:16.82 & -27:20:16.84 &  -12.28 & 0.09 & 27.6$\pm$0.2 & 25.6$\pm$0.2& 1.64$\pm$0.12&  11$\pm$7 & C & 4.70 & -\\
UDG 20 & 10:38:04.43 & -27:29:50.18 &  -12.95 &	 0.14 & 27.3$\pm$0.1 & 26.0$\pm$0.3 & 1.97$\pm$0.12 & - & C & 3.88 & 4693$\pm$12\\
UDG 21 & 10:36:54.17 & -27:36:55.07 &  -12.78 &	 0.11 & 27.3$\pm$0.5 & 24.0$\pm$0.4 & 1.50$\pm$0.12 & - & C & 3.88 & 3489$\pm$22\\
UDG 22 & 10:34:40.97 & -27:42:03.27 &  -13.92 & 0.36 & 26.5$\pm$0.1 & 25.3$\pm$0.2& 3.60$\pm$0.12& - &  P & 1.50 & {  5198$\pm$5}\\
UDG 23 & 10:35:27.70 & -27:46:16.58 &  -14.11& 0.34 & 27.2$\pm$0.2 & 24.3$\pm$0.3 & 2.47$\pm$0.20 & - & C & 2.10 & 3496$\pm$33\\
UDG $32$ & 10:37:04.20& -27:42:53.92&	-14.65& 8.00 & 27.5$\pm$1.0 & 26.2$\pm$1.0& 3.80$\pm$1.00 & 7$\pm$4 & C & 5.00 & -\\
& & & & & & & & & \\
\hline\hline
& & & & & & & & & \\
LSB 1 & 10:36:00.03& -27:28:58.17 &  -12.38&  0.06 & 26.6$\pm$0.1 & 23.9$\pm$0.2 & 0.81$\pm$0.90& - & S & - & - \\
LSB 2 &	10:36:09.65& -27:30:51.61&  -12.87&  0.15 & 25.8$\pm$0.1 & 23.8$\pm$0.1 & 0.57$\pm$0.12& - & P & 0.75 & - \\
LSB 3 &	10:36:26.55 & -27:32:41.61 & -12.03 & 0.05 & 26.6$\pm$0.8 &  23.7$\pm$0.6 & 0.70$\pm$0.12 & - & S & - & - \\
LSB 4 & 10:36:19.71& -27:13:41.68&  -13.83&  0.25 & 26.5$\pm$0.1 & 24.7$\pm$0.1 & 1.48$\pm$0.12 & 8$\pm$3 & C & 3.00 & 3420$\pm$44 \\
LSB 5 &	10:36:41.26& -27:48:20.54&	 -13.74& 0.40 & 25.8$\pm$0.1 & 23.9$\pm$0.1 & 1.42$\pm$0.12 & - & P & 0.75 & {  3439$\pm$40}\\
LSB 6 &	10:38:04.67& -27:32:44.99& -15.47& 1.84 & 26.0$\pm$0.1 & 23.0$\pm$0.2& 4.00$\pm$1.00 & 2$\pm$1 & C & 2.30 & 5193$\pm$5\\
LSB 7 &	10:36:18.70& -27:37:17.93&	 -15.68& 2.59 & 25.3$\pm$0.1 & 22.7$\pm$0.1 & 1.97$\pm$0.10 & - & P & 0.75 & 3551$\pm$11\\
LSB 8 &	10:37:54.47& -27:15:31.12&	 -15.27& 1.44 & 25.1$\pm$0.1 & 23.2$\pm$0.2& 1.51$\pm$0.20& - & P & 0.75 & 2718$\pm$5\\
\hline
\end{tabular}
\tablefoot{Column 1 reports the name of the target in the LEWIS sample. In columns 2 and 3 we list the coordinates. 
In columns 4 and 5 are reported the total $r$-band magnitude and stellar mass, respectively, derived from
the $r$-band image.
Columns 6 to 8  list the structural parameters in the $g$ band published by \citet{Iodice2020b} and \citet{LaMarca2022b}: 
the effective surface brightness, the central surface brightness and the effective radius in kpc, respectively. 
Magnitudes and colours are corrected for Galactic extinction using values from \citet{Schlegel98}.
{  Column 9 reports the total number of GCs which is statistically estimated from photometric data within 1.5 $R_{\rm e}$ and corrected for photometric and spatial incompleteness \citep{Iodice2020b, LaMarca2022b}}. 
Column 10 reports the status of the data acquisition until May 2023, being {C} for {\it completed}, {P} for {\it partially completed}, and {S} for {\it scheduled}. 
The total integration time in hours is listed in Column 11. 
Missing values are for those targets that are either not fully observed yet or are pending final analysis. Their velocities will be reported in a future paper.
In column 12 we report the systemic velocity derived in this work. Missing values are for those targets that are not observed yet.}
\end{table*}

\section{Observations and data reduction}
\label{sec:data}

The LEWIS observations are carried out  
with the MUSE integral-field spectrograph, mounted on the Yepun Unit Telescope 4 at the ESO Very Large Telescope in Chile. 
MUSE is used in Wide Field Mode without adaptive optics, providing a field of view (FoV) of 1$\times$1~arcmin$^2$, 
with a spatial sampling of 0.2$\times$0.2~arcsec$^2$. 
The nominal wavelength range of MUSE is from 4800 to 9300 \AA\ with a spectral resolution (FWHM) that varies
from 2.74 \AA\ (69 km/s) at 5000 \AA\ to 2.54 \AA\ (46 km/s) at 7000 \AA\ \citep{Bacon2017}.

Observations started in December 2021 during ESO period P108, and continued in periods P109 and P110. They are acquired in Service Mode, under dark and clear conditions.
In the first two observing periods (P108 and P109) we gave priority to those UDGs that are fully consistent 
with the empirical classification 
proposed by \citet{vanDokkum2015}, i.e. $R_{\rm e} \geq 1.5$~kpc and $\mu_{0,g} \geq 24$~mag/arcsec$^2$, 
having also photometrically detected GC candidates.
Observations scheduled in P108 and P109 have been completed. The data acquisition for P110 targets, which mostly are LSB galaxies not classified as UDGs, is ongoing,
with a completion of $\sim$20\% in May 2023. 
In each run, and for each target, observations are executed in two steps. Shallower data are acquired
to confirm the redshift first, and, for the confirmed cluster members, we obtained longer exposures to reach the depth required for our scientific purposes. 

Since galaxies in the LEWIS sample span a wide range of effective surface brightness 
$25 \leq \mu_{\rm e} \leq 27$~mag/arcsec$^2$, the
total integration time adopted for each target was set by a required limiting magnitude $\mu_{\rm lim}=\mu_{\rm e}$ and a minimum SNR$=$7 in a spectral bin ($=2.51$\AA) of 2$\times$2 pixels for the brighter targets and 5$\times$5 pixels for the fainter targets. Given that, the total integration times range from 2 hours for galaxies with 
$\mu_{\rm e} \simeq 25$~mag/arcsec$^2$ up to $\sim$6 hours for targets with $\mu_{\rm e} \simeq 27$~mag/arcsec$^2$. 
The total integration time for each target and the status of the data acquisition 
are reported in Table~\ref{tab:UDGsample}. 
Between single exposures we applied a dither of 0.3 to 1.3 arcseconds and a rotation by $90^\circ$, in order to minimise the signature of the 
24 MUSE slices on the field of view. 

\subsection{Data reduction}
Data are initially reduced with the MUSE pipeline version 2.8.5 \citep{Weilbacher2016, Weilbacher2020}, under the ESOREFLEX environment
\citep{Freudling2013}. The main steps include bias and overscan subtraction, flat fielding correction, wavelength calibration, determination of the line spread function, and illumination correction. 
Since all LEWIS targets are less extended than the MUSE FoV (see also Fig.~\ref{fig:UDG11_image}), 
the sky has been evaluated directly on the science frames, as described in the following section.
The flux calibration was obtained using spectro-photometric standard stars observed as part of the MUSE calibration plan. For each galaxy of the sample, the single exposures 
were aligned using reference stars and then combined to produce a first version of a stacked MUSE cube. In this first pipeline reduction, some of the default pipeline parameters for the sky subtraction and alignment for stacking were optimised\footnote{Sky subtraction: \texttt{SkyMethod=auto}, \texttt{skymodel\_ignore=0.02}, \texttt{SkyFr\_2=0.1}; Source alignment: \texttt{threshold=8}, \texttt{bkgfraction=0.2}, \texttt{srcmin=8}}.

\subsection{Sky subtraction}\label{sec:skysub}
After running the standard MUSE data reduction workflow, we used the resulting data products to obtain the final data cube with an improved sky subtraction, adapting the workflow described in \citet{Zoutendijk2020}. 
The method and tools described below have been applied to one UDG of the sample, UDG11, chosen as a test case 
(see also Sec.~\ref{sec:data_quality}). 
We used the reconstructed $r$-band image from the first version of the stacked MUSE cube 
to create an object mask. Objects were detected on the background-subtracted and Gaussian kernel convolved $r$-band image with the \texttt{photutils} software \citep{larry_bradley_2020_4044744}. 
The resulting object mask was dilated by a factor of two and covers all visually detected foreground and background objects in the $r$-band image. To also cover the faint outskirts of the UDG that remained unmasked by the automatic object detection, we manually added an ellipse covering the central $1\;R_{\rm e}$ of UDG~11 \citep[$R_{\rm e} = 1.66\,\mathrm{kpc}$, ellipticity $\epsilon=0.3$, position angle P.A.$=144\degr$;][]{Iodice2020b}.

We then ran ESOREFLEX again with the following modifications to the standard workflow. As we now used custom sky masks, we increased the fraction of pixels considered as the sky to \texttt{SkyFr\_2=0.75}. 
Masking any non-sky objects in the sky also allowed us to use the autocalibration routines and we thus set \texttt{autocalib=deepfield}. The autocalibration method was originally developed for the MUSE deep field and uses the sky background to estimate correction factors for each slice in several wavelength bins after the rejection of outliers \citep{Weilbacher2020}, removing the spatial structure from MUSE exposures that remained after flat fielding. We also applied multiplicative flux calibration to each exposure before exposure combination, accounting for varying sky backgrounds\footnote{This workflow is not ingested into the ESOREFLEX pipeline, we acknowledge here a private communication with Lodovico Coccato at ESO}. 
After applying these factors to the pixtables of the individual exposures, we combined them with the standard ESOREFLEX workflow. 

Additional cleaning of the residual sky contamination was performed on the final stacked datacube, 
using the Zurich Atmospheric Purge algorithm \citep[ZAP;][]{Soto2016}. 
ZAP was run within the ESOREFLEX environment, where a dedicated workflow is available within the MUSE pipeline. 
We parsed the custom sky mask to the workflow and tested different parameter combinations to obtain the 
optimum sky subtraction results. 
The reconstructed images of the final sky-subtracted data cube for UDG11, 
obtained with the standard prescriptions for the sky-subtraction and with the improved method described above,
are shown in Fig.~\ref{fig:UDG11_image}. 
The reconstructed images are displayed with the same intensity levels, which emphasises the improvement in the background level residual noise in our adopted “improved data reduction” 
(see top left panel of Fig.~\ref{fig:UDG11_image}), with respect to the cube reduced with the standard sky-subtraction technique (top right panel of Fig.~\ref{fig:UDG11_image}).


\section{Cluster membership}\label{sec:redshift}

The first goal of the LEWIS project is to confirm the cluster membership of all targets in the sample.
To this aim, for each galaxy, we have derived the systemic velocity ($V_{\rm sys}$) by fitting the stacked 
spectrum inside $1R_{\rm e}$ with the Penalised Pixel-Fitting code \citep[pPXF;][]{Cappellari2017}, 
using the MUSE rest-frame wavelength range between 4800 and 7000 \AA. 
During this step, the region of the spectrum at longer wavelengths is excluded, since they are strongly affected 
by residuals from the sky line subtraction. 
Before obtaining the stacked spectrum, we used the MUSE reconstructed image to mask 
all the background and foreground bright sources. In addition, in the stacked spectrum are also 
masked all those wavelength regions with strong sky residuals.
The MUSE reconstructed images and the stacked spectra, for all so far 
observed LEWIS galaxies, are shown in the Appendix~\ref{sec:images}.

We have used the E-MILES single stellar population (SSP) models \citep{Vazdekis2012, Vazdekis2015} as 
spectral templates.
They have a spectral resolution of FWHM~$= 2.51$ \AA\ \citep{FalconBarroso2011} and cover a large range 
in age (from 30 Myr to 14 Gyr) and total metallicity ($-2.27 \leq$ [$M$/H] $\leq 0.4$ dex). 
The error estimate for each value of $V_{\rm sys}$ corresponds to the formal error provided by pPXF.
The $V_{\rm sys}$ value for each of the observed galaxies is reported in Table~\ref{tab:UDGsample}.

{  The distribution of $V_{\rm sys}$ values is plotted in Fig.~\ref{fig:red_hist}. 
It ranges from 2317 km/s to 5198 km/s, with a peak value around $\sim 3500$~km/s, which coincides 
with the peak of the velocity distribution for the Hydra I bright ($m_B\leq16$~mag) cluster members and dwarf galaxy
population.
The mean cluster velocity, $V_{\rm sys} = 3683 \pm 46$~km/s \citep{Christlein2003}, 
and its velocity dispersion, $\sigma_{\rm cluster}\sim700$~km/s \citep{Lima-dias2021}, 
can be used to determine the membership of our targets: 14 of the 20 galaxies with the measured velocities
are cluster members because they are found to have their velocity within the cluster velocity dispersion
($3438 \leq V_{\rm sys} \leq 4302$~km/s). Since the cluster is quite isolated in the recession velocity space between 
2000 and 5000\,km/s \citep{Richter1982,Richter1987}, and the velocity distribution of its galaxy members is broad, 
the three galaxies with velocities inside $2\sigma_{\rm cluster}$ are also likely cluster members (UDG4, UDG8 and UDG12, see Table~\ref{tab:UDGsample}).The three remaining galaxies, UDG22 with $V_{\rm sys} = 5198 \pm 5$~km/s, 
LSB6 with $V_{\rm sys} = 5193 \pm 5$~km/s, 
and LSB8 with $V_{\rm sys} = 2718 \pm 5$~km/s,
could be background or foreground galaxies. 

%

In the photometric work of \citet{Iodice2020b} and \citet{LaMarca2022b}, to discriminate
UDGs from normal dwarfs LSB galaxies via their physical sizes, we assumed that all 
newly detected galaxies are at the distance of the Hydra I cluster (i.e. 
51 Mpc). Now, having confirmed the cluster membership of most UDGs and LSB galaxies
with our LEWIS spectroscopy, we stick to the fixed distance for deriving 
their effective radii and confirming their morphological classification, 
as already published in our previous papers (see also Table~\ref{tab:UDGsample}). We are aware
that the Hydra I cluster might have a physical depth of a few Mpc, and thus
some of the UDGs might be located in front or back of the cluster, resulting
in size differences of $\pm 4$\% for a relative distance of 2 Mpc with respect to Hydra I.

For the three outliers in the redshift distribution, we have used 
the Hubble law, assuming H$_0$=70 km/s/Mpc, to derive the Hubble flow distance based 
on the measured $V_{\rm sys}$. 
For UDG22 and LSB6, we obtained a distance of 74 Mpc and the new
values for their effective radius are $R_{\rm e}=5.22$~kpc and $R_{\rm e}=5.80$~kpc,
respectively. For LSB8, which has a lower $V_{\rm sys}$, $R_{\rm e}=1.12$~kpc.
Therefore, UDG22 and LSB6 turn out to be more diffuse and fainter, if they 
would be a UDG and a LSB in the background of Hydra I cluster.
LSB8 is smaller and, according to its central surface brightness $\mu_{0,g}=23.20$~mag/arcsec$^2$, it can be classified as a foreground dwarf galaxy.
}


\section{Data quality and preliminary results: UDG11 as test case}\label{sec:data_quality}

This section describes the analysis of the MUSE data for one of the sample galaxies, UDG11, 
to assess the data quality. UDG11 has been chosen as a test case since it is one of the faintest 
and most diffuse galaxies in the sample, with a comparably large number of photometrically detected 
GC candidates (see Table~\ref{tab:UDGsample}), 
for which we obtained all the requested data at the desired depth.
With this target, we customize and test the data reduction process as well as analysis tools and methods that will later be applied to the entire LEWIS sample. 
In the following sections, we describe
the analysis of the MUSE cube for UDG11 to derive the {\it i)} line-of-sight velocity distribution LOSVD,
{\it ii)} stellar population properties, and {\it iii)} systemic velocities of the GC candidates.

UDG11 is located on the west side of the cluster (see Fig.~\ref{fig:Hydra}). 
The structural properties for this galaxy, based on deep images, were published by \citet{Iodice2020b}.
In detail, UDG11 has an absolute magnitude $M_{r}=-14.75$~mag in the $r$ band and a 
stellar mass  $M_{\star}=0.63 \times 10^8$~M$_\odot$.
The structural parameters derived from the 1D fit of the azimuthally-averaged surface brightness profiles, 
in the $g$ band, are $\mu_{0,g}=24.36\pm0.13$~mag/arcsec$^2$ and $R_{\rm e}=1.66\pm0.12$~kpc.
The integrated colour is $g-r=0.43\pm0.11$~mag. {  Based on the colour selection and shape, the estimated total number of GCs\footnote{  Here and throughout the paper, $N_{\rm GC}$ is the total number of the globular clusters hosted in the galaxy obtained after  photometric and spatial incompleteness correction, as described in Sec. 2.1 of \citet{Iodice2020b}.} in this galaxy is $N_{\rm GC}=7\pm3$ \citep[see][and Table~\ref{tab:UDGsample}]{Iodice2020b}.}

\begin{figure*}
\centering
\includegraphics[width=9cm]{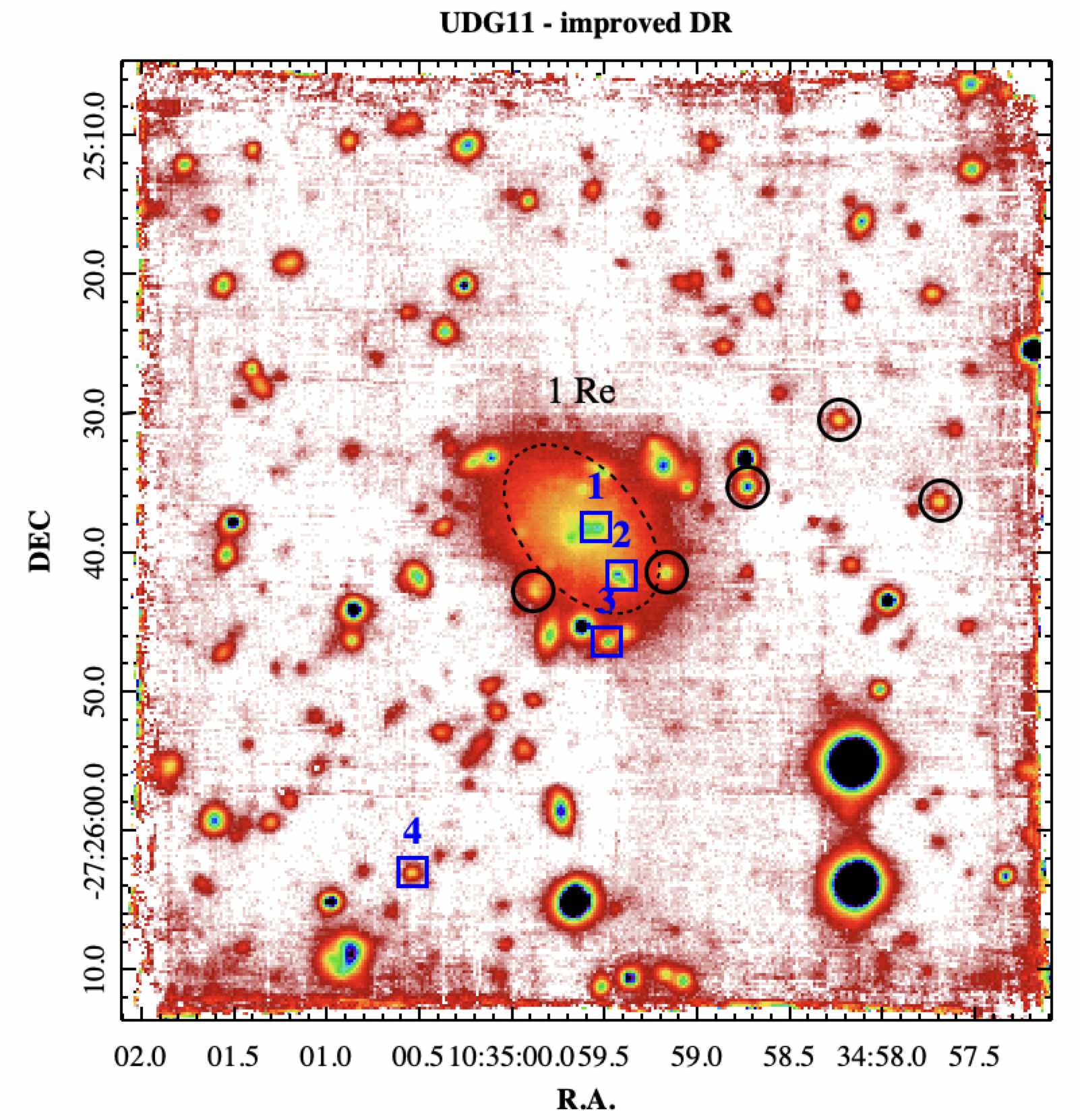}
\includegraphics[width=9cm]{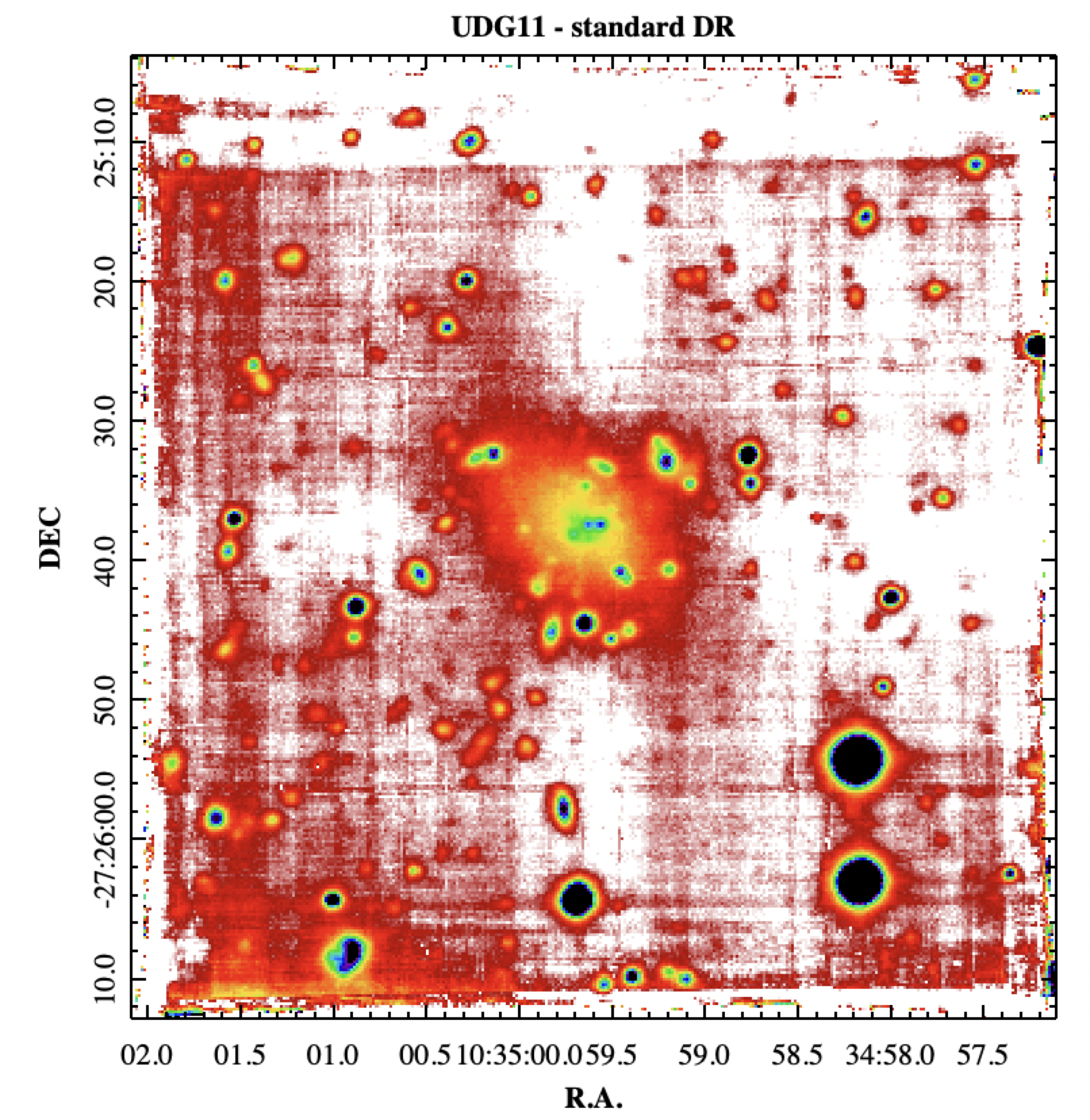}
\includegraphics[width=9cm]{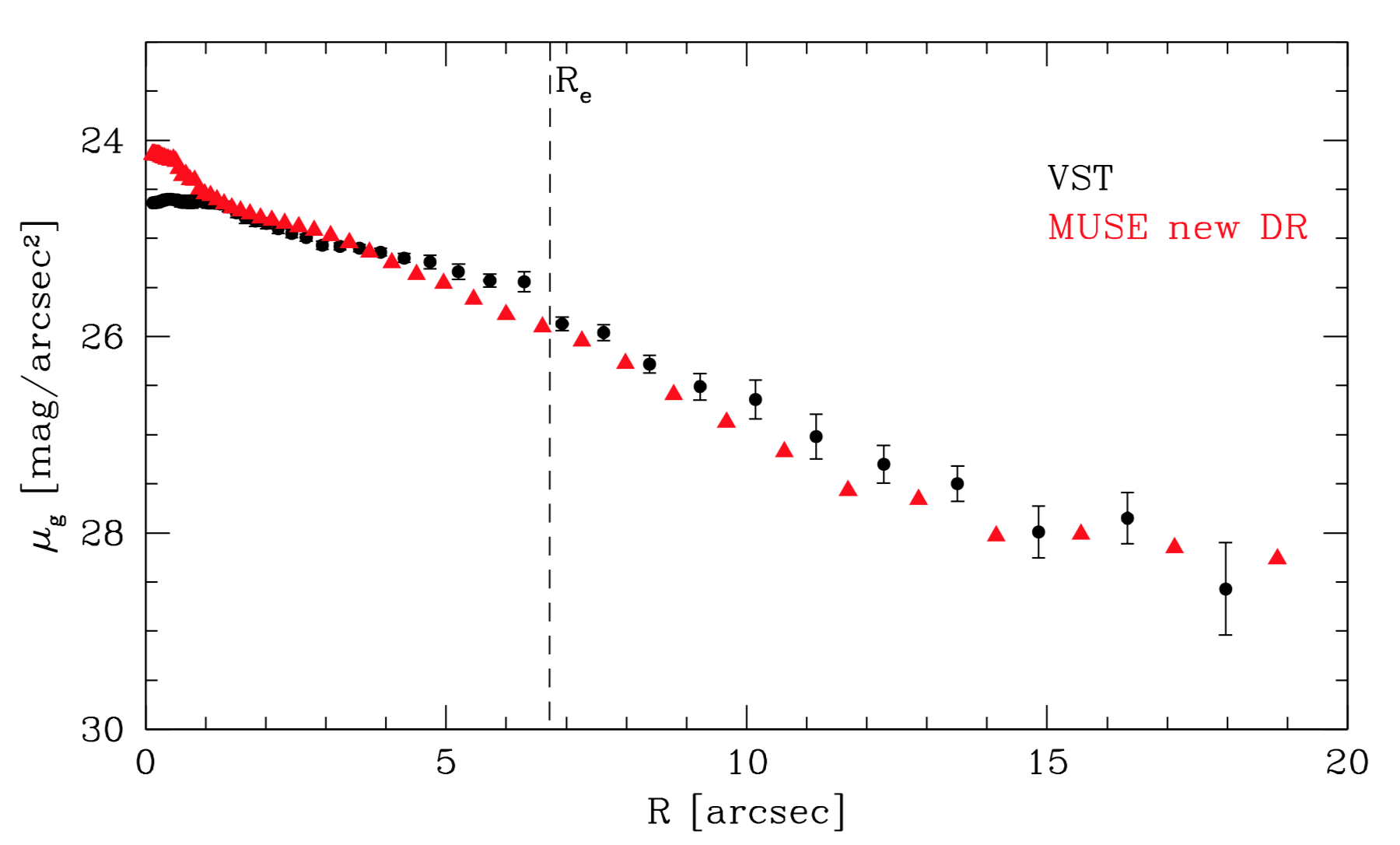}
\includegraphics[width=9cm]{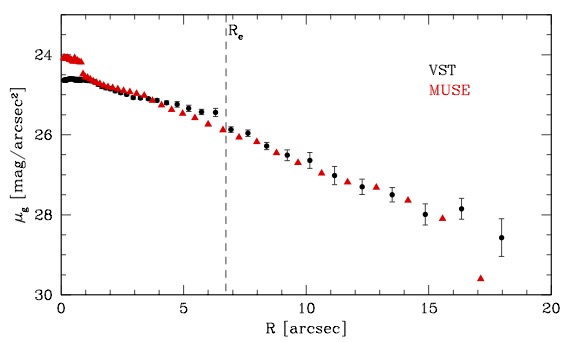}
\caption{{  MUSE reconstructed image of UDG11. Top panels - reconstructed images obtained by improving the sky subtraction (left panel), as described in Sec.~\ref{sec:skysub}, and with the standard
 prescriptions for the data reduction (right panel). 
 On the left panel, the dashed ellipse marks the isophote at $1R_{\rm e}$.
 The confirmed GCs are marked with the blue numbered boxes, see also Table~\ref{tab:GC_prop}. GCs 1 and 3 have radial velocities consistent with UDG11, while GCs 2 and 4 have radial velocities inconsistent with UDG11 but still consistent with the Hydra~I cluster. The photometrically pre-selected GC candidates from \citet{Iodice2020b} are marked with black circles. Four of them were 
 found to be emission line galaxies, and for the other one it was not possible to retrieve the $V_{\rm sys}$ due to the low SNR (see Sec.~\ref{sec:GCs} for details).}
 Bottom panels - Azimuthally averaged surface brightness distribution derived from the MUSE reconstructed image (red triangles) compared with the same profile derived from the VST optical $g$-band image (black circles), obtained by improving the sky subtraction (left panel), and with the standard
 prescriptions for the data reduction (right panel).  In both panels, the difference between VST and MUSE surface brightness profiles inside $\sim$1~arcsec is due to the seeing-limited observations of the optical VST images.}
\label{fig:UDG11_image}
\end{figure*}

\begin{figure}
    \centering
    \includegraphics[width=\hsize]{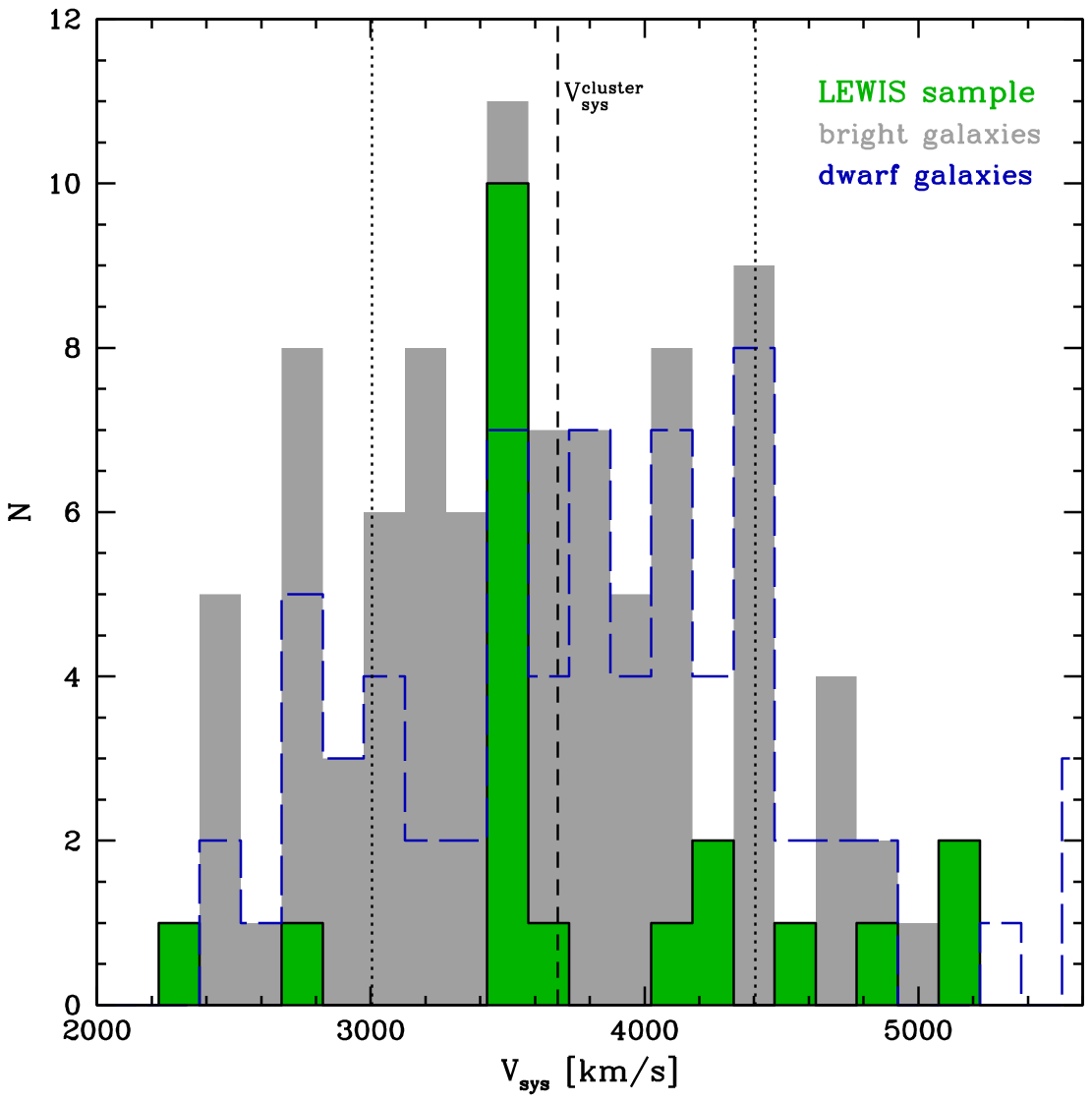}
    \caption{  Velocity distribution of the galaxies in the LEWIS sample (green histogram). 
    The distribution of velocities for the bright cluster members ($m_B<16$~mag) and for the dwarf galaxy population in the Hydra I cluster are shown with the grey and long-dashed blue histograms, respectively, from the catalogue by \citet{Christlein2003}.
    The vertical dashed line marks the cluster mean velocity of $3683 \pm 46$~km/s. The vertical dotted lines correspond to the average velocity dispersion of the cluster members $\sim 700$~km/s \citep{Lima-dias2021}.}
    \label{fig:red_hist}
\end{figure}

The MUSE cube obtained for UDG11 has a total integration time of 6.10 hours. The MUSE reconstructed 
images (from the whole wavelength range), resulting from the standard data reduction and improved procedure (see Sec.~\ref{sec:skysub}), 
are shown in Fig.~\ref{fig:UDG11_image}. In the lower panels we compare, after arbitrarily 
re-scaling, the azimuthally-averaged surface brightness profiles from the MUSE reconstructed images with
that of the optical VST image in the $g$ band. 
This illustrates that with the MUSE data for UDG11 we are able 
to map the integrated light down to $\mu_g\sim 28$~mag/arcsec$^2$ and out to $\sim$2$R_{\rm e}$, 
suggesting also that a satisfactory level of sky subtraction has been achieved.
With the improved data reduction the residual patterns of the MUSE slices are well removed, 
even if the continuum appears slightly over-subtracted when compared to photometric profiles
(Fig.~\ref{fig:UDG11_image}).

\begin{figure*}
	\centering
	\includegraphics[width=1\hsize]{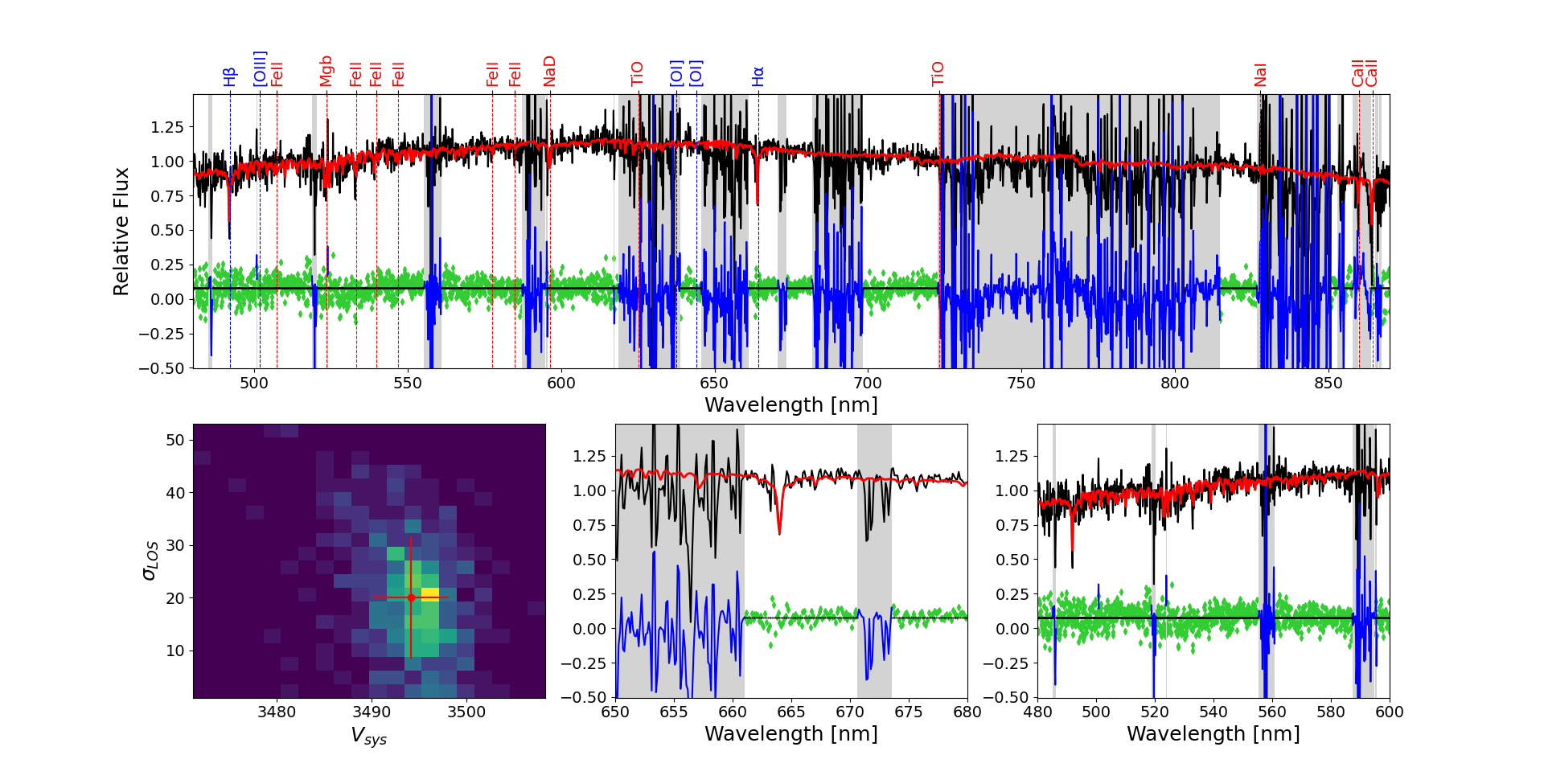}
	\caption{  Restframe stacked MUSE spectrum (black solid line) for UDG11 derived inside $1R_{\rm e}$. The main absorption features are marked as dashed red (absorption features) and blue lines (absorption features, but potentially also in emission) and labels. The red solid line represents the best fit obtained with pPXF. Green points are the residuals between the observed spectrum and its best fit. The grey areas are the masked regions of the spectra, excluded from the fit. Bottom (right and middle) panels show the enlarged regions around the main fitted absorption features of the top-row fit in the optical wavelength range $4800-7000$~\AA. The lower-left panel shows the grid of values for the line-of-sight velocity ($V_{\rm LOS}$) and velocity dispersion ($\sigma_{\rm LOS}$) derived from the Monte Carlo simulations (see text for details).}
	\label{fig:UDG11_1Respectra}
\end{figure*}

From the MUSE cube obtained with the improved data reduction, and used for the analysis described in this paper, we have extracted the stacked spectrum
inside a circular area with a radius $R=1 R_{\rm e}$, where all bright sources (background galaxies and 
foreground stars) are masked. This is shown in Fig.~\ref{fig:UDG11_1Respectra}.
It is worth noting that the SNR per spaxel of the stacked spectrum from 
this cube is SNR=16, which is higher than 
the SNR=12.6 we computed for the spectrum from the standard data reduction. In addition, 
by masking the wavelength regions affected by the sky-lines residuals, the SNR increases to 20.

In this spectrum, we can clearly identify the absorption features of the most relevant
lines, as H$\beta$, Mgb, H$\alpha$ and the CaT (see also lower boxes in Fig.~\ref{fig:UDG11_1Respectra}). On the other hand, we do not detect emission lines, suggesting that this galaxy is devoid of ionised gas.


\subsection{Stellar kinematics}\label{sec:kin}


The stellar kinematics of UDG11 have been derived by using the pPXF code. 
Specifically, we extract the LOSVD, which is parametrised by the line-of-sight (LOS) velocity $V$, 
velocity dispersion $\sigma$, and Gauss-Hermite moments $h_3$ and $h_4$ \citep{Gerhard1993,vanDerMarel1993}. 
We have used the E-MILES stellar library (\citealp{Vazdekis2016}) with single stellar population (SSP) models as spectral templates. These models have a spectral resolution comparable 
to the average of the MUSE spectra, i.e. $\sim$2.5 \AA\,(see Sec.~\ref{sec:data}) which makes them suitable for our purposes. 
For all models, we have assumed a Kroupa IMF \citep{Kroupa2001}.

To assess the effectiveness of the MUSE cube and identify any potential constraints in measuring stellar kinematics caused by the low-surface brightness levels of these galaxies, we examined the entire MUSE rest-frame wavelength range, from 4800 to 9000 \AA, as well as a restricted range, from 4800 to 7000 \AA, which is less susceptible to the residual effects of sky line subtraction.

Since a reliable extraction of the higher-order Gauss-Hermite 
moments requires a relatively high SNR \citep[e.g.][see also Fig.~\ref{fig:test_sigma} of this
study for exact estimations]{Gadotti2005},
for UDG11 we have derived only the first 
two moments of the LOSVD, i.e. $V$ and $\sigma$.
These quantities have been estimated by running pPXF to fit the 1D stacked spectrum inside $1R_{\rm e}$.
On the full MUSE rest-frame wavelength range between 4800 and 9000~\AA, the best fit was obtained by allowing high-order multiplicative and additive 
Legendre polynomials of degree 10. Including additive polynomials allows to correct for differences 
in the flux calibration and overcome the limitations of the stellar library, whereas multiplicative polynomials directly alleviate imperfections in the spectral calibration affecting the continuum shape. 
Results are shown in Fig.~\ref{fig:UDG11_1Respectra}. 
A similar approach was already used for the UDG NGC1052-DF2 by \citet{Emsellem2019} to estimate the stellar kinematics from MUSE data.
The mean LOS velocity (which we adopted as systemic velocity) and velocity dispersion of UDG11 are 
$V_{\rm sys}=3507 \pm 3$~km/s and $\langle \sigma \rangle_{\rm e} = 20 \pm 8$~km/s, respectively.
Restricting the fit to the range 4800$-$7000 \AA, 
the best fit provides $V_{\rm sys}=3509 \pm 3$~km/s and 
$\langle \sigma \rangle_{\rm e} = 20 \pm 10$~km/s, being consistent with the former values.

The error estimates on $V_{\rm sys}$ and $\langle \sigma \rangle_{\rm e}$ are obtained by performing Monte Carlo simulations \citep[e.g.][]{Cappellari2004,Wegner2012}. This procedure is described in Appendix~\ref{sec:errors}.

\begin{figure}
	\centering
\includegraphics[width=\hsize]{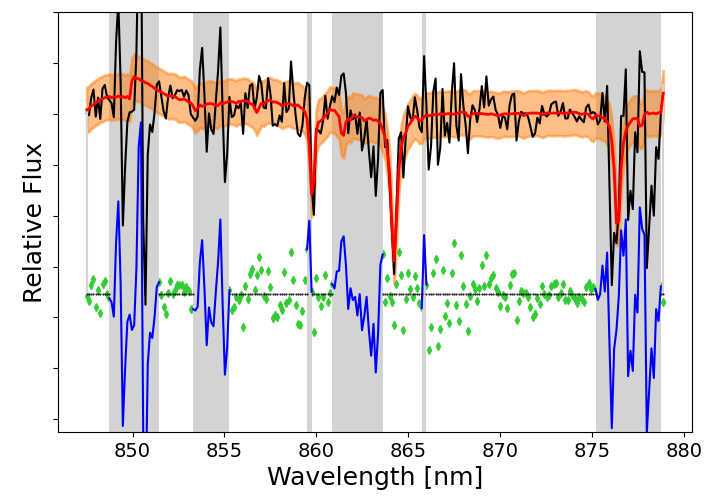}
	\caption{Stacked MUSE spectrum for UDG11 in the CaT region, inside $1R_{\rm e}$. The red solid line represents the best fit obtained with pPXF. Green points are the residuals between the observed spectrum and its best fit. The grey areas are the masked regions of the spectrum, excluded from the fit. The orange region corresponds to the standard deviation of the best fit across all the $\sim1400$ iterations.}
	\label{fig:CaT_fit}
\end{figure}

The stacked spectrum for UDG11 shows prominent absorption CaT lines (see Fig.~\ref{fig:CaT_fit}).
These lines are the strongest features in the stellar continuum for a large variety
of stellar types \citep{Cenarro2001}, in addition to being at the MUSE wavelength region with better resolution \citep{Bacon2017}, they are particularly suitable to extract 
the stellar kinematics. We run pPXF on the wavelength region restricted to the rest-frame interval 
8475$-$8690 \AA.
To fit exclusively the CaT region, we used the CaT templates by \citet{Cenarro2001}. Given that the spectral resolution
of these templates is $\sim$1.5\,\AA, they have been convolved to the resolution of the MUSE 
spectra in this range ($\sim$2.5\,\AA). The best fit, shown in Fig.~\ref{fig:CaT_fit}, 
has been obtained with multiplicative and additive Legendre polynomials of degree 7, 
and by masking the reddest CaT lines, which is strongly affected by a residual of the sky lines.
We obtain $V_{\rm sys}=3501 \pm 5$~km/s and $\langle \sigma \rangle_{\rm e}=16 \pm 12$~km/s.
These values are consistent with the previous estimates, provided above, even if $\langle \sigma \rangle_{\rm e}$ has a larger error. All quantities, computed with the three different methods, are
reported in Tab.~\ref{tab:UDG11}.

\subsection{  Tests on the velocity dispersion measurements}\label{sec:sigma_tests}
\begin{figure*}
    \centering
    \includegraphics[width=\hsize]{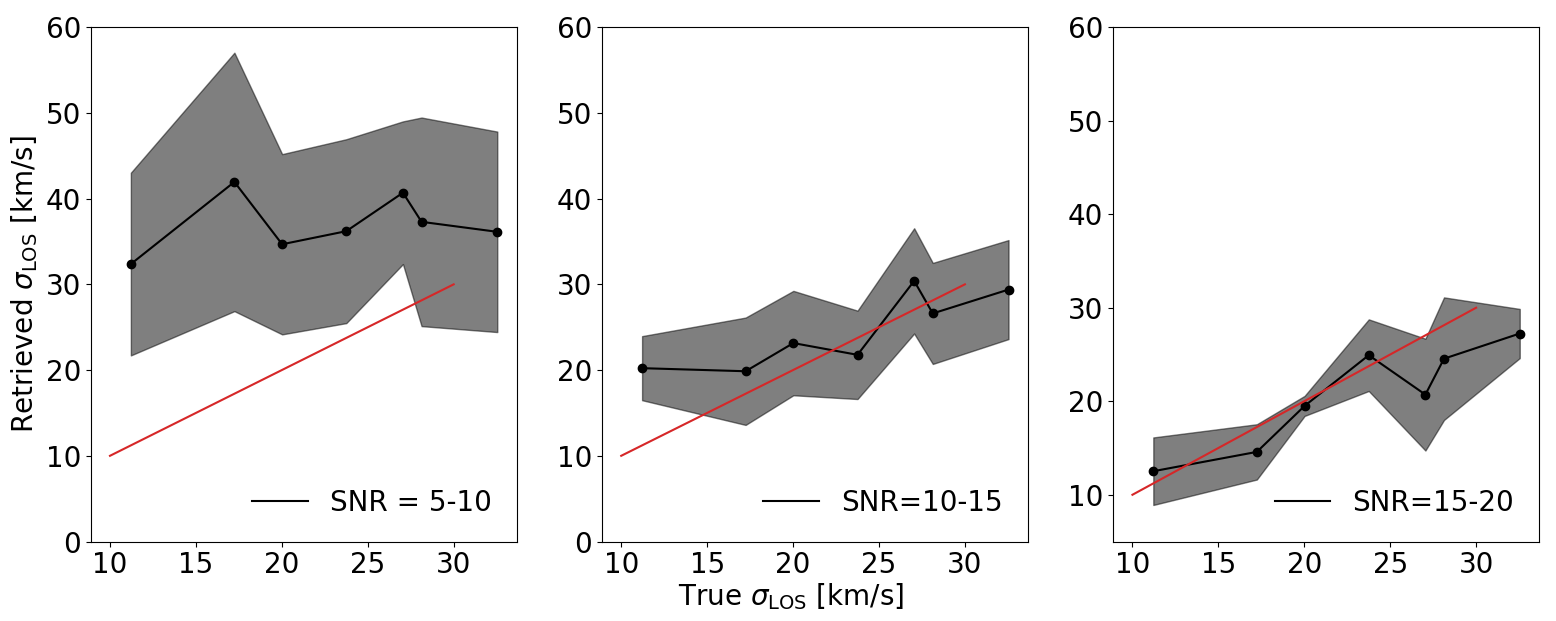}
    \caption{Results of simulations testing the ability of our pipeline to successfully recover a velocity dispersion of different values at low signal-to-noise (per pixel) levels, from SNR=5-10 (left panel), SNR=10-15 (middle panel), up to SNR=15-20 (right panel). Shaded areas correspond to the standard deviation of the measured values at every SNR. The red solid line corresponds to the output values equal to the input ones (unity line).
    In this experiment, the error is dominated by Poisson noise and no systematic sources of uncertainties have been considered. See Appendix~\ref{sec:poisson_tests} for details. }
    \label{fig:test_sigma}
\end{figure*}

{ 
By fitting the stacked spectrum inside 1$R_{\rm e}$, we have derived a very low value for $\langle\sigma\rangle_{\rm e} \sim 20$~km/s, less than the MUSE spectral resolution (see Sec.~\ref{sec:data}).
As stated in the previous section, the fit has been performed by using the latest version of the pPXF code, developed by \citet{Cappellari2017}. 
The author proved that, for spectra with high SNR (SNR > 3000 per spectral element), the full spectrum fitting provides reliable kinematics at any velocity dispersion, even below the instrumental resolution of MUSE.

Using the SAMI integral-field spectrograph to study the dwarf galaxies in the Fornax clusters, \citet{Eftekhari2022} 
proved that a velocity dispersion of $\sim0.4$ the instrumental 
resolution can be measured, at SNR=15, with a $\sim 20-30$\% of accuracy.

The LEWIS data are in the low-surface brightness regime, therefore, 
the main issue is the SNR of the spectra. 
Given that, we have performed several tests to check which is the minimum 
SNR needed for the data 
to retrieve a reliable value of $\sigma_{\rm LOS}$. 
We simulated mock spectra based on the E-MILES models, with different SNR ranging from 
5 to 120 by introducing 
Poissonian noise. A full description of this test is reported in Appendix~\ref{sec:poisson_tests}.
We found that, from spectra with SNR~$\sim 15-20$ (comparable to that of the stacked spectrum of UDG11),
we can retrieve velocity dispersion as low as $\sigma_{\rm LOS} \sim 10$~km/s with an uncertainty of 10 km/s. Results are shown in Fig.~\ref{fig:test_sigma}.
Similar tests were performed by \citet{Eftekhari2022}, and we found consistent results.

In conclusion, our tests demonstrate that we need a minimum SNR per spaxel of 10 for an unbiased velocity determination. Given that the stacked spectrum of UDG11, has SNR=16, the best-fit value for its velocity dispersion is $\langle\sigma\rangle_{\rm e}=20\pm8$~km/s. This might be considered an upper limit, because we cannot exclude lower values.

In addition to the tests described above, we have performed a totally independent check of what is 
the expected value of the velocity dispersion for UDG11.
By adopting the scaling relation derived by \citet{Zaritsky2023}, 
where the total mass-to-light ratio $\Gamma_{\rm e}$ derived inside $R_{\rm e}$ ($\Gamma_{\rm e}$) is a 
function of the velocity dispersion and the effective luminosity $I_{\rm e}$, we derived $\sigma = 19.97$~km/s.  In this relation, we have used $L_{\rm e}=3.35\times 10^7$~L$_\odot$, derived by 
\citet{Iodice2020b}, and we have assumed a total $M/L=10$, constant with radius,
typical for dwarf galaxies of luminosity comparable to UDG11 \citep[e.g.][]{Battaglia2022}.
The resulting value of the velocity dispersion is fully consistent with the estimate obtained
by fitting the MUSE cube.
Therefore, even considering the limits of the data described above, we 
conclude that the derived estimate of the velocity dispersion for UDG11 
is very close to the expected value for this galaxy.}

\subsection{Spatially-resolved stellar kinematics maps}\label{sec:2Dkin}
\begin{figure*}
\centering
\includegraphics[width=\hsize]{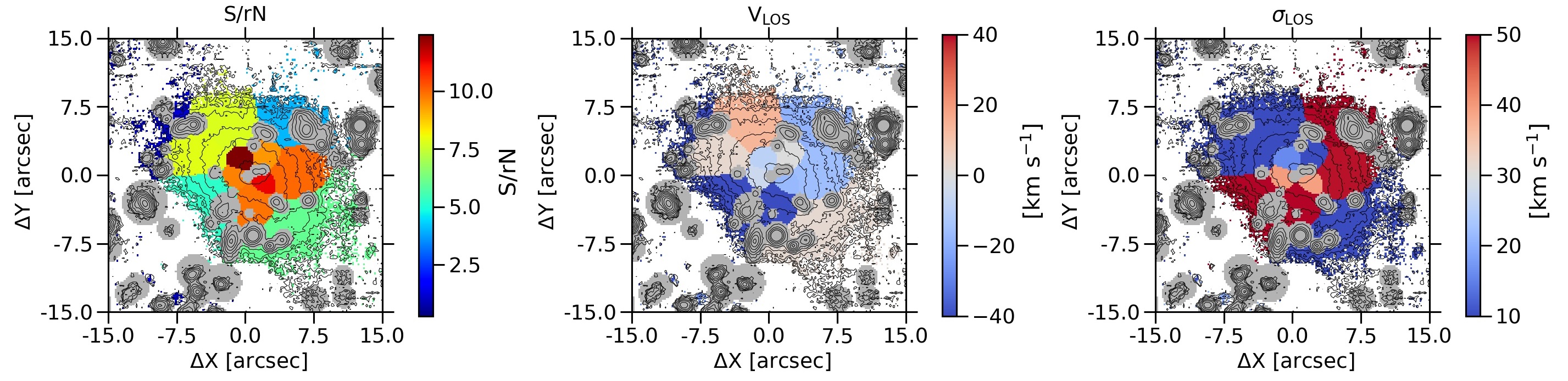}
\caption{2-dimensional kinematic maps of UDG11: {  signal-to-residual noise ratio S/rN} (left), LOS velocity $V_{\rm LOS}$ (centre) and velocity dispersion $\sigma_{\rm LOS}$ (right) derived from a Voronoi binning with SNR=10. Contours correspond to the isophotes of the light distribution. Grey circles represent the adopted mask. The FoV is 30\,$\times$\,30 arcsec and is oriented North up and East left.}
\label{fig:2Dmaps}
\end{figure*}

The 2D map of the stellar kinematics is derived by using the modular Galaxy IFU Spectroscopy Tool (GIST) pipeline
for the analysis of IF spectroscopic data, developed by \citet[][]{Bittner2019}. 
Using GIST, we spatially binned the datacube spaxels with the adaptive algorithm by \cite{Cappellari2003} based
on Voronoi tessellation.
Since the average SNR of the stacked spectrum is $\sim$16, as binning threshold we adopted SNR$=$10 (Fig.~\ref{fig:2Dmaps}, left panel). 
The fit was restricted to 
the optical wavelength range $4800-6800$ \AA \ to avoid the region of the CaT, which is affected by the 
sky lines residuals, particularly in the galaxy outskirts. 
As starting guesses, we have used the same ``set-up'' of the best fit of the 1D stacked spectrum
inside $1 R_{\rm e}$. Therefore, we adopted the E-MILES stellar templates, 
additive polynomials of grade 10, and multiplicative polynomials of grade 10. 
In Figure~\ref{fig:2Dmaps} we show the map of the Voronoi bins, with the average signal-to-residual S/rN (left panel),
the resulting 2D maps of the LOS velocity $V_{\rm LOS}$ (middle panel) and 
velocity dispersion $\sigma_{\rm LOS}$ (right panel).
The SNR map shows that all bins close to the centre of the galaxy, corresponding to the brightest 
regions, have SNR$\geq10$ per spaxel, which is the targeted threshold fixed for the Voronoi tessellation.
In the galaxy outskirts, some bins have SNR$\leq 10$ per spaxel. This effect was discussed by 
\citet[][see Fig.6 of that paper and references therein]{Sarzi2018} for the MUSE cubes.
They found that the quality of the Voronoi-binned spectra decreases with the surface brightness,
where lower values of the SNR with respect to the formal threshold chosen for the Voronoi bins are found at the lowest surface brightness levels. This is due to the impact of the spatial 
correlations between adjacent bins at lower surface brightness levels.

\begin{figure}
\centering
\includegraphics[width=9cm]{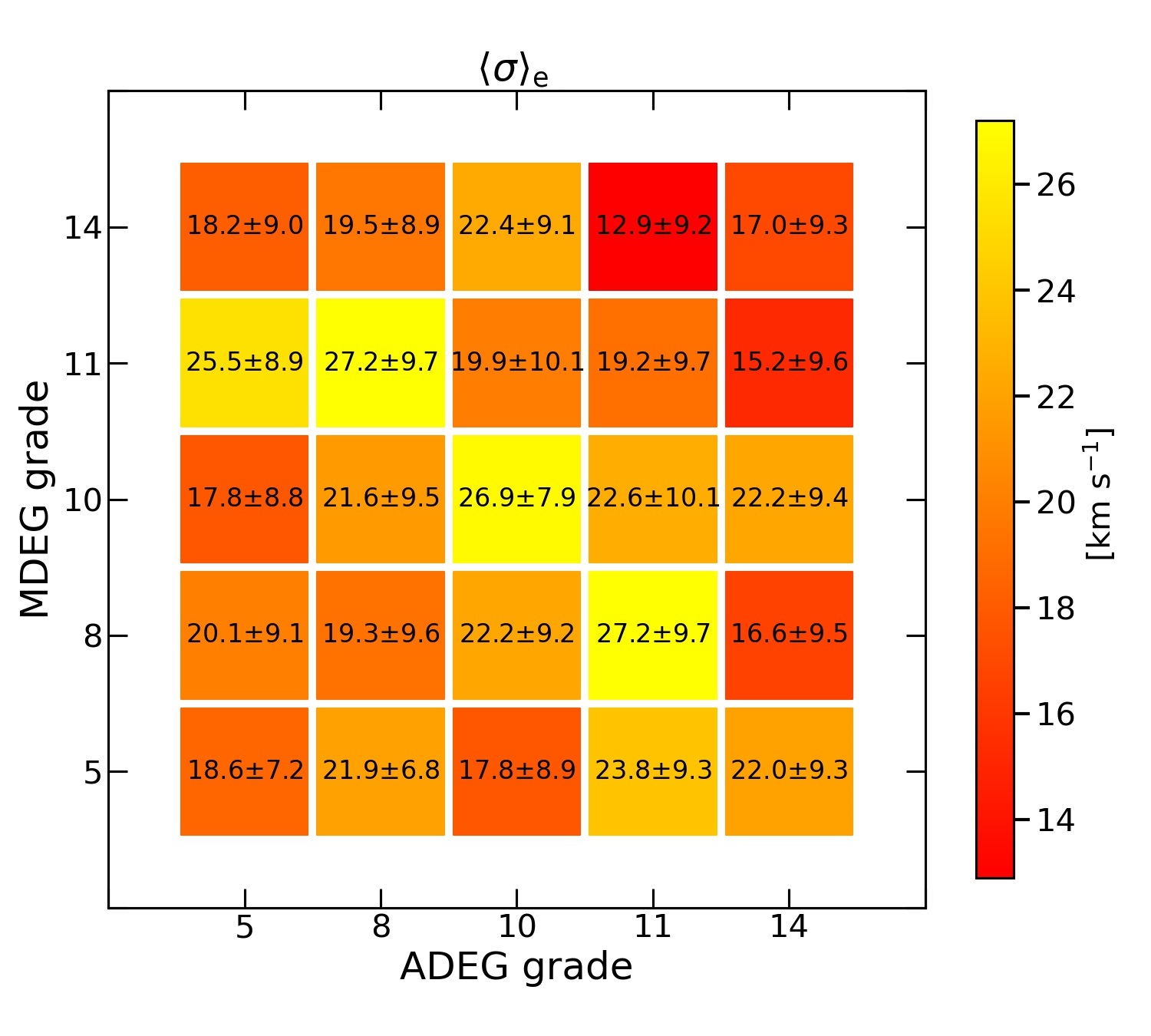}
\caption{  Values of the weighted mean of $\langle\sigma \rangle_{\rm e}$ derived from the fit of the Voronoi binned cube (with SNR=10 per spaxel) for different degrees of multiplicative and additive Legendre polynomials.
Boxes are colour-coded according to the value of $\langle\sigma \rangle_{\rm e}$. See text for details.}
\label{fig:test_poly_sigma}
\end{figure}

From the 2D map of the LOS velocity dispersion (Figure~\ref{fig:2Dmaps}, right panel), 
we extracted the effective velocity dispersion $\langle \sigma \rangle_{\rm e}$ by calculating 
the weighted mean of the values enclosed in an elliptical region with a semi-major axis equal to the 
effective radius and with axial ratio $q = 1-\epsilon$, with ellipticity $\epsilon=0.3$. 
The associated error was estimated by calculating the standard error of the weighted mean. We found $\langle \sigma \rangle_{\rm e} = 27 \pm 8$\,km/s, which is consistent within $1\sigma$ error with 
the different estimates derived from the 1D spectrum (Table~\ref{tab:UDG11}).
However, we have also tested more options where the degree of Legendre polynomials spans a larger
range of values from 5 to 14. The resulting values of $\langle \sigma \rangle_{\rm e}$ are
shown in Fig.~\ref{fig:test_poly_sigma}. 
In order to be consistent with 
the stellar kinematics derived from the 1D stacked spectrum, we adopted the 2D maps obtained with
the same multiplicative and additive polynomials (Fig.~\ref{fig:2Dmaps}).

The 2D $V_{\rm LOS}$ map does not show a clear trend of rotation along any direction, the 2D $\sigma_{\rm LOS}$ map shows values ranging from $\sigma_{\rm LOS} \sim 10$~km/s to $\sigma_{\rm LOS} \sim 30$~km/s along the major axis,
and larger values of $\sigma_{\rm LOS} \sim 40-50$~km/s along the minor axis of the galaxy.  
We checked the reliability of these measurements 
by inspecting the fit of the spectrum in each bin. We found that some bins in the outskirts have a SNR lower with respect to the adopted threshold for the Voronoi binning, therefore, in these bins, the values of
$\sigma_{\rm LOS}$ are affected by a systematic overestimation, as addressed in Sec.~\ref{sec:kin}
(see Fig.~\ref{fig:test_sigma}).


\subsection{Stellar populations}\label{sec:sfh}

For UDG11, we have performed an additional pPXF run on the stacked spectrum inside $1 R_{\rm e}$ in order to 
 derive its mean age and total metallicity. We have fitted the full MUSE rest-frame 
 wavelength range between 4800 and 9000 \AA, using the E-MILES templates 
 (Padova isochrones with Kroupa IMF; \citealp{Girardi2000}). 
As first step, we have generated a reference fit, to be used also as a central setting for the 
generation of bootstrapped spectra and as reference for Monte Carlo iterations over various 
parameters (see Appendix~\ref{sec:errors}). This step involves the creation of a mask, 
flagging all noisy pixels and sky lines affecting the spectra (with a flux greater than three 
times the noise level). 
The process also includes setting polynomial degrees as shown in Sec.~\ref{sec:kin}: 
we utilise the same framework as the 1D stacked spectrum, but with the exclusion of 
additive polynomials. In addition, we also fix our previously found 1D kinematic 
solutions to those from the stellar population fits, constraining in this way the velocity dispersion.

To construct the final solution, pPXF uses a technique called regularization, which allows the routine to preferentially select the smoothest solution among many (products of the well-known age-metallicity degeneracy; see e.g, \citealp{Cappellari2017}). 

\begin{figure*}
\centering
\includegraphics[width=9.15cm]{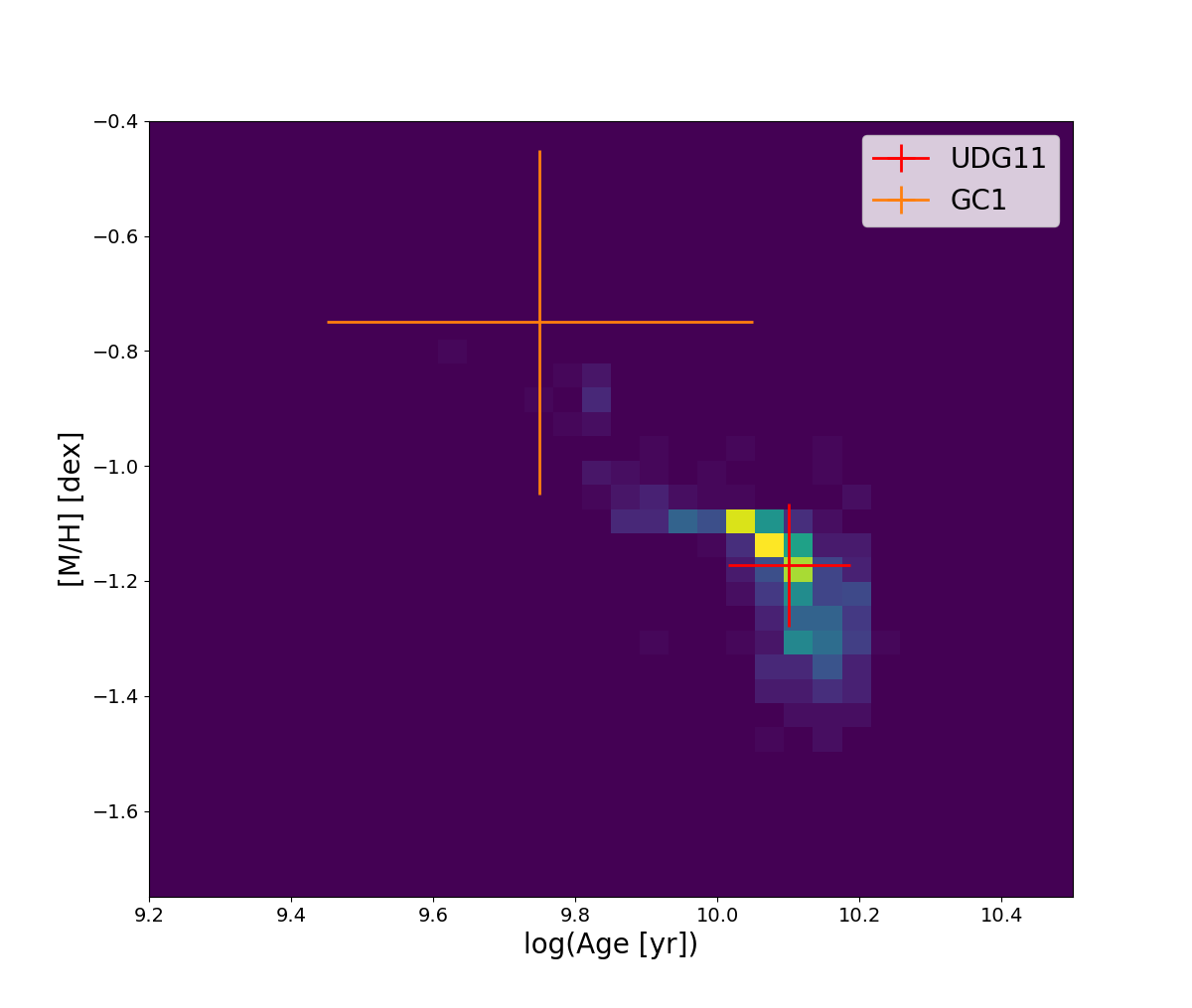}
\includegraphics[width=9cm]{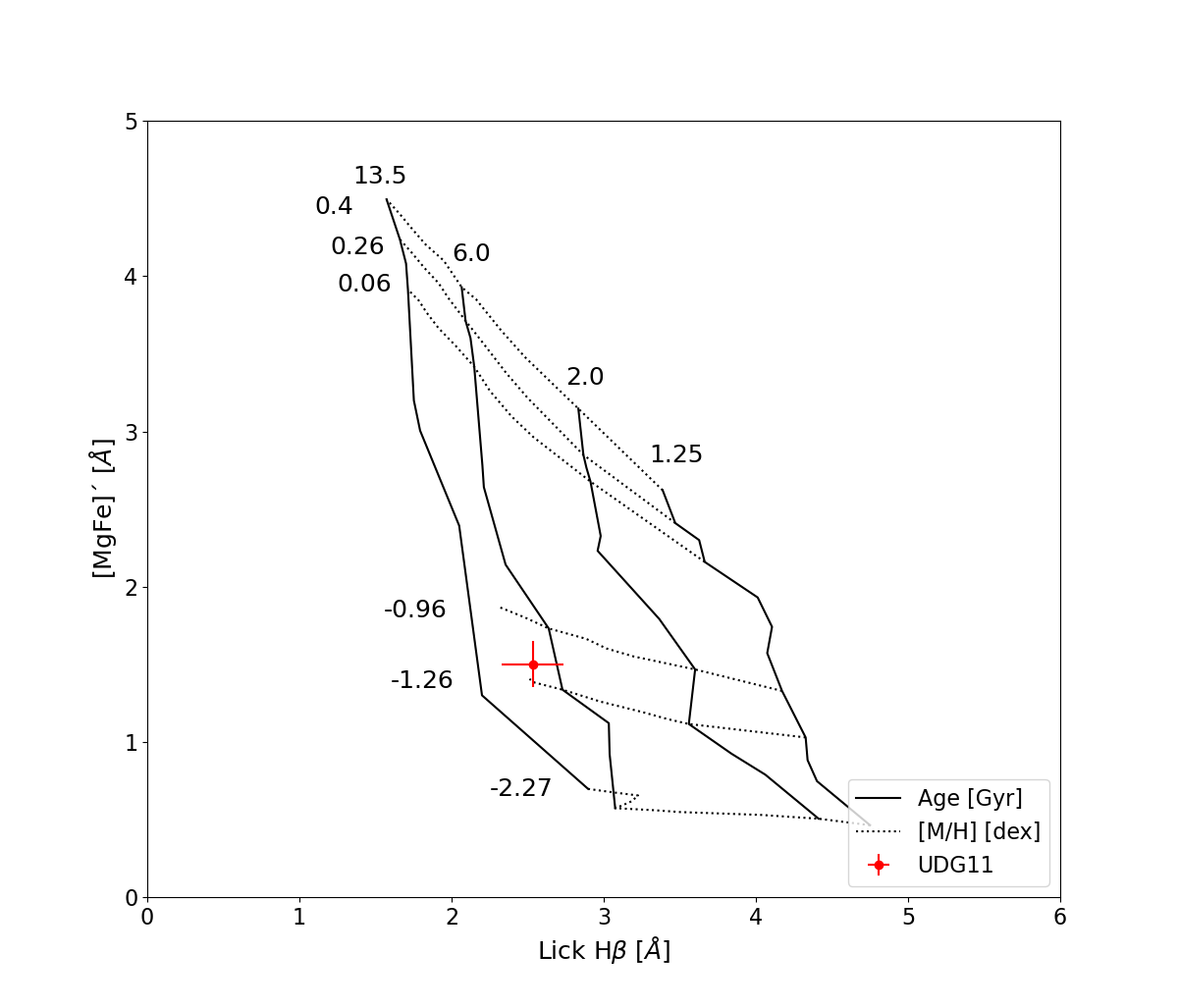}
\caption{Mean age and metallicity for UDG11 derived from the stacked spectrum inside $1R_{\rm e}$. 
 Left panel - Best fit derived using the pPXF code.
 The red cross corresponds to the values for age and metallicity (including errors) obtained by the best fit on the whole MUSE spectral range. The orange cross shows the average age and metallicity (including errors) for one of the confirmed GCs in UDG11 (see Sec.~\ref{sec:GCs}). 
 Right panel - Lick indices measurements. Using the E-MILES BaSTI templates, we constructed the grid: vertical dashed lines denote models for the same age (left being the oldest), and horizontal dotted lines refer to model measurements for spectra with the same metallicity (with the most metal poor ones located at the lower part).}
	\label{fig:sfh}
\end{figure*}

Over 700 iterations, the median age and metallicity which we 
obtained by fitting the whole MUSE 
wavelength range, are Age$ = 10 \pm 1$~Gyr and [$M$/H]$ = -1.17 \pm 0.11$~dex, respectively.
%
By fitting the stacked spectrum for one of the confirmed GCs (see Sec.~\ref{sec:GCs}), 
we obtained comparable age ($7.9\pm1.5$~Gyr) and metallicity ([$M$/H]$ = -0.7 \pm 0.3$~dex). Results are plotted in Fig.~\ref{fig:sfh} (left panel) and listed in Table~\ref{tab:UDG11}.

As a self-consistency check, we take the E-MILES set of templates (BaSTI isochrones with the same Kroupa IMF) and compute the 
Lick H$\beta$ and [MgFe]' indices of various templates in order to construct a reference grid. 
This computation is done with the latest version of \emph{pyphot}\footnote{\url{https://mfouesneau.github.io/pyphot/}} without following the usual Lick indices convention (see e.g, \citealp{Vazdekis2010}) for the assumed FWHM. Instead, to avoid losing potentially useful information, we keep the resolution of the data intact, as it roughly matches the one of the templates used to construct the grid.
To compute the errors on the Lick indices, we take Lick measurements for every best fit model from 
each converging pPXF iteration, and then use the 1$\sigma$ measurement on the Monte Carlo distribution 
to estimate the error. We find the Lick indices measurements to be fairly consistent (Lick H$\beta = 2.54 \pm 0.2$ and [MgFe]' $= 1.49 \pm 0.1$)
with the stellar population measurements of age and metallicity.
Results are shown in Fig.~\ref{fig:sfh} (right panel).

\begin{table*}[]
    \centering
        \caption{Stellar kinematics, age and metallicity derived for UDG11 from the LEWIS data.}
        \label{tab:UDG11}
    \begin{tabular}{lccccc}
    \hline
    \hline
    \smallskip
     Fit & $\lambda$ range & $V_{\rm sys}$ & $\langle\sigma\rangle_{\rm e}$  & [$M$/H] & Age\\
         & [\AA]           & [km/s]        & [km/s]                    & [dex]   & [Gyr]\\
       (1) & (2) & (3) & (4) & (5) & (6)\\
        \hline
      1D & 4800-9000   & 3507$\pm$3  & 20$\pm$8    &  $-1.2\pm$0.1 & 10$\pm$1\\
      1D & 8475-8690   & 3501$\pm$5  & 16$\pm$12 & -- & -- \\
      2D & 4800-7000   & 3532$\pm$43 & 27$\pm$8   &  -- & -- \\
      GC1 & 4800-9000  & 3503$\pm$12 & --          & $-0.7\pm$0.3 & 9$\pm$2\\
      \hline
    \end{tabular}
    \tablefoot{In col.1 and col.2 we report the kind of fit we have performed or source we have used and the wavelength range used to derive the stellar kinematics and age and metallicity, respectively. In col. 3 and 4 are listed the systemic velocity and velocity dispersion derived for UDG11 inside $1R_{\rm e}$ (see Sec.~\ref{sec:kin}). In col. 5 and 6 are reported the estimates of the metallicity and age derived for UDG11 inside $1R_{\rm e}$ (see Sec.~\ref{sec:sfh}), and for GC1.}
\end{table*}


\subsection{Globular cluster population}\label{sec:GCs}

From the UDG11 MUSE cube, we extracted and analysed the spectra of the photometrically 
pre-selected GC candidates 
from \citet{Iodice2020b}, as well as all other point-like sources within the luminosity range 
expected for GCs (see also Fig.~\ref{fig:UDG11_image}). 
Where the signal-to-noise was sufficient, we measured the radial velocity 
to verify their kinematical association with the UDG. 

We only considered the sources with SNR~$\ge 2.5$ 
in the background subtracted spectrum, using an 8-pixel circular aperture. 
For lower SNR, the $V_{\rm sys}$ could not be reliably determined. 
This SNR limit corresponds to an apparent magnitude of $m_g \sim 25.5$~mag, 
about 0.5 mag brighter than the expected turn-over
magnitude (TOM) of the GC luminosity function (GCLF) at the distance of the Hydra I cluster \citep{Iodice2020b}.
 
For measuring the radial velocity of the candidate GCs, we used the SSP model spectra from the E-MILES\footnote{We restricted the library to ages $\ge$ 8 Gyr. According to \citet{Fahrion2020}, this choice appears reasonable because most GCs have ages $>$8 Gyr with only very rare exceptions.} library to fit each GC spectrum
with pPXF. In this case, we used an IMF with a double power-law (bimodal) 
shape and a high mass slope of 1.30 \citep{Vazdekis1996}. 
We did not attempt to measure the intrinsic velocity dispersion of the candidate GCs, due to the low spectral resolution of MUSE and the limited SNR of the observed spectra (see also Fig. \ref{fig:test_sigma}). 

\begin{figure*}
    \centering
    \includegraphics[width=\hsize]{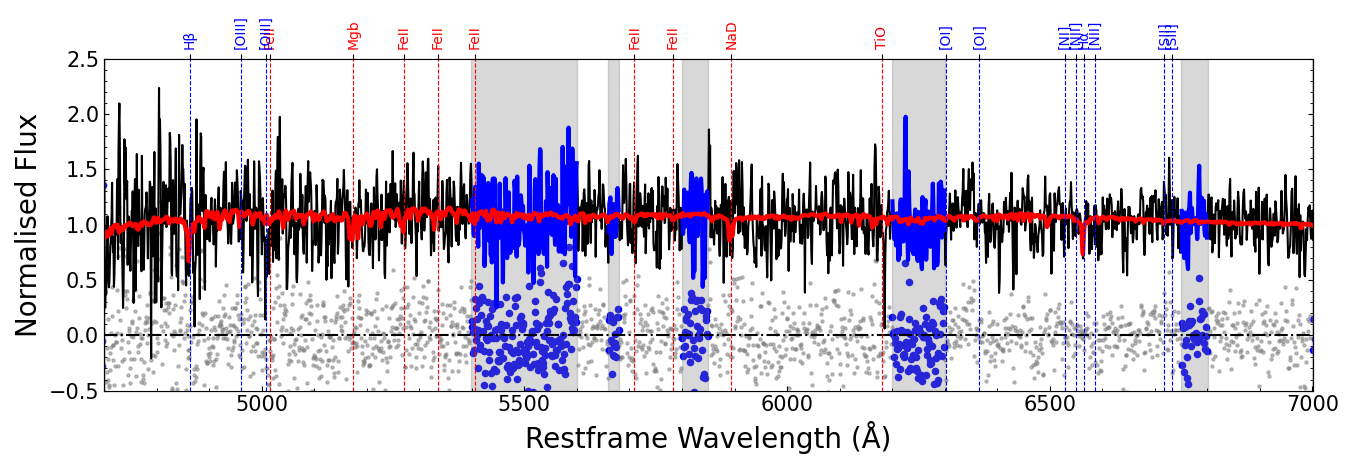}
    \caption{Stacked spectrum of GC3, with SNR $\sim$4.5 \AA$^{-1}$. The original spectrum is shown in black, the pPXF fit in red. Regions with strong sky residual lines were masked from the fit (grey-shaded areas). }
    \label{fig:GC_spectra}
\end{figure*}

Figure \ref{fig:GC_spectra} shows the spectrum of one GC in the field, and the corresponding pPXF 
fit as an example. The regions excluded from the spectrum are grey-shaded and  
correspond to the residual sky or telluric lines. Moreover, we  excluded from the fit 
the wavelength region with $\lambda \geq 7000 $\AA\
 because of the presence of strong sky residuals
which made the identification of the CaT not feasible. 

Also in this case, to get a reliable uncertainty on the estimate of $V_{\rm sys}$ we fitted the sources in a Monte Carlo technique approach (see Sec.~\ref{sec:kin}). 
After the first fit using the original spectrum, we created 200 realizations with the same SNR of the 
original spectrum by perturbing the noise-free best-fit spectrum\footnote{Where by 'best-fit spectrum' we refer to the 'best fitting template spectrum'.} with random draws in each wavelength bin from the residual (best-fit subtracted from the original spectrum). The fit was then repeated and the LOS velocity
was determined from the mean of the resulting distribution. The random uncertainty is given by the standard 
deviation assuming a Gaussian distribution. 
A detailed description of the method to study the GC population in UDGs, including UDG11, 
is the subject of a forthcoming paper by Mirabile et al. (in prep.).

We identified four point sources in the UDG11 cube with a $V_{\rm sys}$ velocity consistent with the 
Hydra\,I cluster. They are listed in Table~\ref{tab:GC_prop} and marked in Figure~\ref{fig:UDG11_image}. The projected position of GC1, GC2 and GC3 is within $2R_{\rm e}$ of UDG11, while GC4 is much more distant ($\sim4R_{\rm e}$). 

The radial velocity of GC1 and GC3 are 3503 $\pm 12$ km/s and 3460$\pm 21$ km/s, respectively.
Compared to the systemic velocity of the UDG11, they are
within $1\sigma$ and $\sim$2$\sigma$, respectively. 
We therefore consider that both GCs are associated to UDG11. 
Moreover, it's worth noting that the velocity differences $\Delta V$ of GC1 and GC3, of
$\Delta V \sim -4\pm13$~km/s and $\Delta V \sim -47\pm22$~km/s (see Table~\ref{tab:GC_prop}), respectively,
are consistent with the stellar velocity dispersion obtained for UDG11 close to the centre ($\sigma_{\rm LOS}\sim20$~km/s) and in the South-East region of the galaxy ($\sigma_{\rm LOS}\sim20-40$~km/s), as shown in Fig.~\ref{fig:2Dmaps}.  

The velocity difference $\Delta V\geq130$ km/s for GC2 and GC4 might suggest they 
are not gravitationally bound to UDG11, and, therefore, they might be 
considered as intra-cluster GCs.
However, this result can only be confirmed once the DM 
content of this galaxy gets better constrained. 
{  To this aim, we calculated the escape velocity assuming a NFW halo profile \citep{Miller2016}, 
and we estimated the DM halo mass using the \citet{Burkert2020} relation. 
We found that, if the DM halo mass lies between 4 and $6\times 10^{10}$~M$_\odot$, the GCs with an 
escape velocity $\leq 200$~km/s can be considered as bound systems. Therefore, GC2 with $\Delta V \simeq260$~km/s should be 
considered an intra-cluster GC, 
whereas all the other GCs could potentially be bound to the UDG11. Finally, the very central GC, GC1, might be a nuclear star cluster.

In summary, the preliminary analysis of the GC systems in UDG11 suggests that this galaxy 
has either two or three spectroscopically confirmed GCs that are bound to the host.
By excluding GC1, as a potential nuclear star cluster and GC2 as an intra-cluster GC, with the remaining two spectroscopically confirmed 
GCs we can estimate the GC specific frequency ($S_N$) of  UDG11.
Our spectroscopic completeness limit is $m_g \sim 25.5$ mag ($m_r\sim24.9 $\,mag), which is 0.5 brighter than the GCLF turnover magnitude of $m_g^{\rm TOM}$=26\,mag \citep{Iodice2020b}. Assuming a GCLF width of $\sigma_{\rm GCLF}$=1\,mag \citep{Villegas2010}, our spectroscopic completeness limit thus corresponds to 34\% of the GCLF. 
The incompleteness corrected total number of GCs then is 2/0.34=$5.9^{+2.2}_{-1.8}$, where the error range comes from the 68\% confidence interval of a Poisson distribution centred on 3.9 (i.e. the number of 2 spectroscopically confirmed GCs has no errorbar, only the incompleteness correction of 3.9 has it). 
The absolute $V$-band magnitude of UDG11 is $M_{V,0}=-14.6 \pm 0.1$ mag. 
Therfeore, we obtain $S_N= 8.4^{+3.2}_{-2.7}$. This value of $S_N$ is
consistent with the typical estimates for dwarf galaxies of
similar luminosity \citep{Georgiev2010,Lim2018}.
By including both GC1 and GC2, the value of $S_N$ would correspondingly change, but it is still consistent within the uncertainties with the values quoted above.}



{  It is worth noting that none of the five photometrically pre-selected GC candidates from \citet[][see Fig.~\ref{fig:UDG11_image}, black circles]{Iodice2020b}, can be confirmed as GC associated to UDG11: four of them exhibit emission lines typical of background galaxies and one has too low SNR for a reliable velocity measurements.
From \citet{Iodice2020b} the number of contaminants is $3.3 \pm 0.8$~arcmin$^{-2}$, therefore, for a region of $5R_{\rm e}$ around the UDG11, we expect a number of contaminants equal to $3 \pm 1$. 
Among the five photometrically pre-selected GCs, one would thus have expected about two spectroscopically confirmed GCs. This is indeed the case. We note that the majority of photometrically pre-selected candidates have not been confirmed, while the two newly detected GCs within the MUSE cube did not pass the narrow photo- and morpho-metric selections adopted for the VST dataset. 
This is not surprising, given the availability of only two photometric and very close optical bands. The preliminary results on other LEWIS targets reveal that UDG11 is rather peculiar in this respect. As an example, out of five photometric preselected GC candidates of UDG3, three are spectroscopically confirmed GCs, whereas the remaining two have too small SNR to provide useful constraints (further details will be presented in Mirabile et al., in prep.).}

\begin{table*}[]
\renewcommand{\tabcolsep}{0.15cm}
    \centering
        \caption{Identified GCs in the UDG11.}
        \label{tab:GC_prop}
    \begin{tabular}{ccccccccc}
    \hline
    \hline
     GC & RA & DEC & $V_{\rm sys}$ & $\Delta V$ & $m_r$ & SNR & Galactocentric Distance & Classification \\
        &[deg]&[deg]&[km/s]&[km/s]&[mag]& &[kpc]&\\
        (1)&(2)&(3)&(4)&(5)&(6)&(7)&(8)&(9)\\
        \hline
        
        1&158.7482&-27.427& 3503$\pm$ 12&  -4$\pm$ 13&24.57 $\pm$ 0.07
        &10.2&<$R_{\rm e}$&GC/Nucleus\\
        2&158.7477&-27.4280& 3767$\pm$ 54&  +260$\pm$ 54&24.05$\pm$ 0.07& 5.8&<$R_{\rm e}$&Intra-cluster GC?\\
        3& 158.7479&-27.4293& 3460$\pm$ 21&  -47$\pm$ 22& 24.18$\pm$ 0.08&4.5 & $<1-2R{\rm e}$& GC\\
        
        4& 158.7523&-27.4339&3640$\pm$ 33&  +133$\pm$ 34&24.88 $\pm$ 0.03& 2.7&$\sim 4R_{\rm e}$&Intra-cluster GC?\\

        \hline
        \hline
    \end{tabular}
    \tablefoot{Col.1: GC number as marked in Fig. \ref{fig:UDG11_image}. Col.2 and Col.3: coordinate of the sources. Col.4: estimated line-of-sight velocity and uncertainty. Col.5: difference between the estimated radial velocity of the sources and the radial velocity of the UDG11. {We adopted as radial velocity for the UDG11 the value reported in Sec. \ref{sec:data_quality}.} Col.6: magnitude in the r-band
    from the VST photometry. Col.7: SNR per pixel of the background-subtracted spectrum. Col.8: Distance of the sources in terms of the effective radius of the UDG.
    Col.9: source classification.}

\end{table*}

\section{The structure of UDG11 from LEWIS data}\label{sec:disc}

In this section, we discuss the stellar kinematics, stellar populations 
and DM content of UDG11, as a result of the analysis of the MUSE cube presented in this paper.
The main goal is to probe the ability of the LEWIS data to constrain these quantities, the relative 
uncertainties, and, therefore, the formation history of UDGs.


\subsection{Stellar kinematics and populations of UDG11}
In Table~\ref{tab:UDG11} we summarised the measurements of the
stellar kinematics (i.e. systemic velocity and velocity dispersion) and 
the average age and metallicity, derived for UDG11 (see Sec.~\ref{sec:kin} and Sec.~\ref{sec:sfh}).
Based on the fit of the 1D stacked spectrum inside $1 R_{\rm e}$,
these estimates suggest that the UDG11 has a very low velocity dispersion
($\langle \sigma \rangle_{\rm e}\sim20$ km/s), it is old ($\sim 10$ Gyr) and metal-poor ([$M$/H]$ \sim -1.2$~dex).
%
The spatially-resolved map of the LOS velocity (see Fig.~\ref{fig:2Dmaps}, central panel), 
does not reveal a significant velocity gradient or rotation along the photometric axes of the galaxy.



\begin{figure*}
    \centering
    \includegraphics[width=9cm]{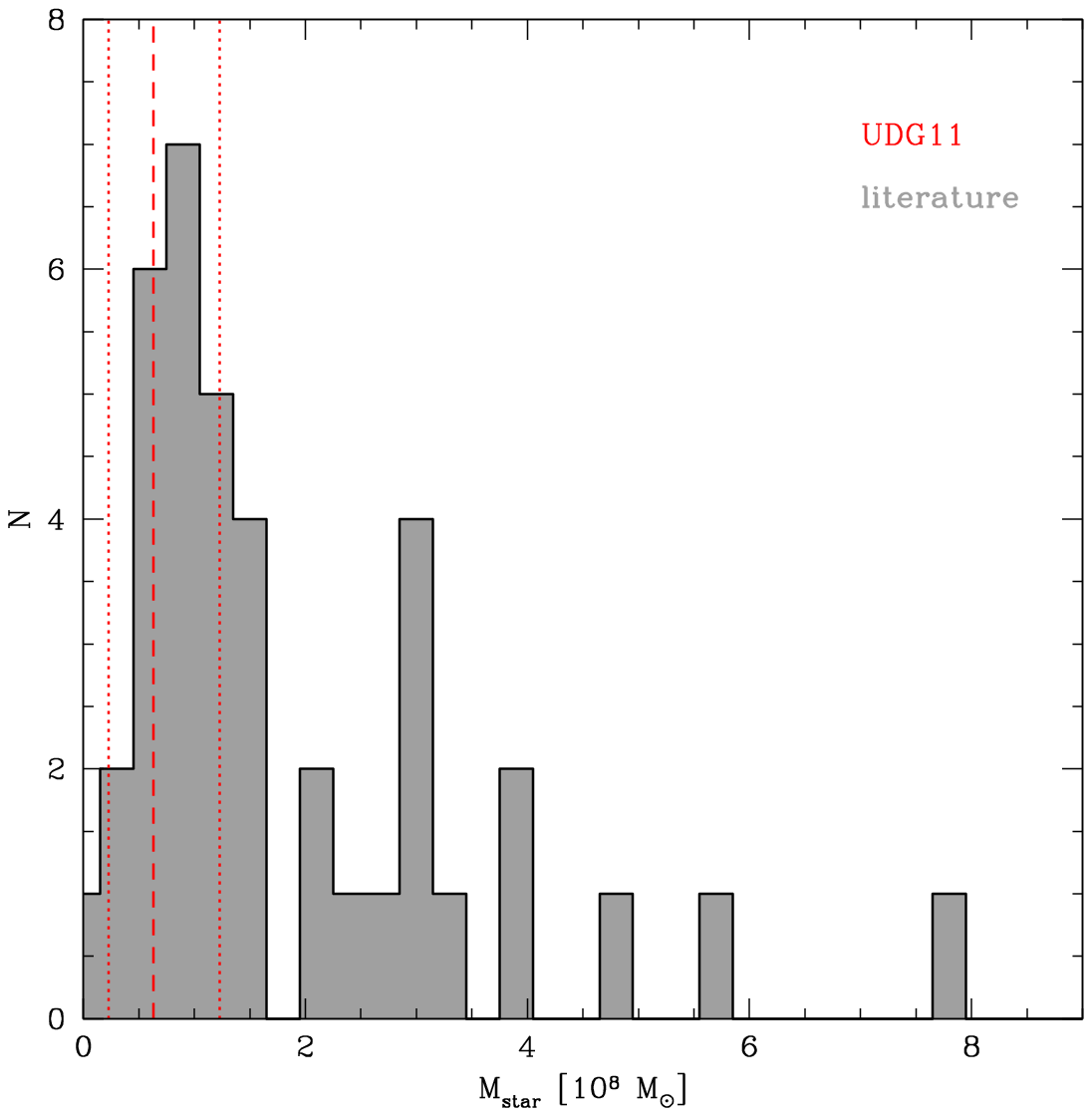}
    \includegraphics[width=9cm]{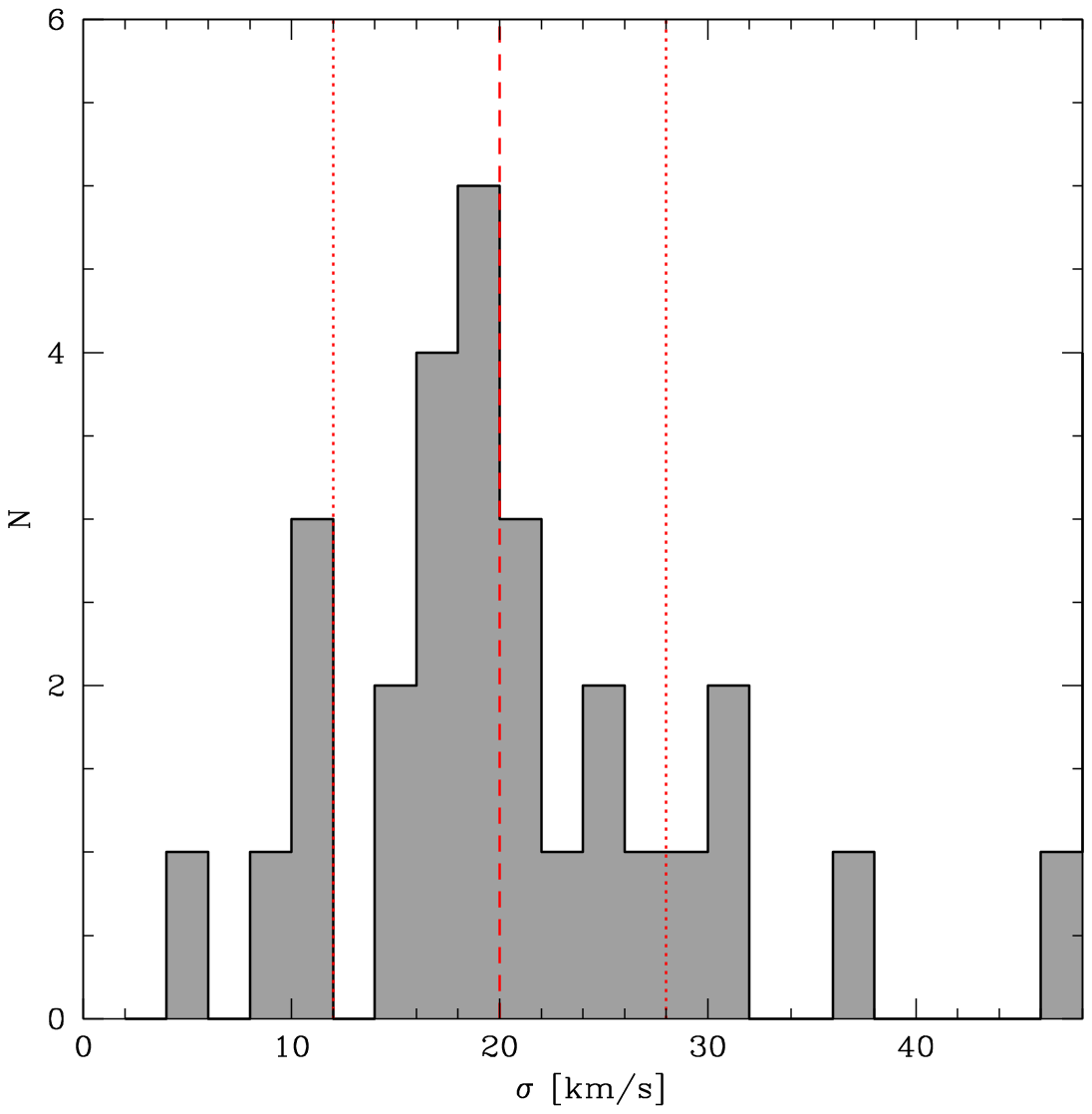}
    \includegraphics[width=9cm]{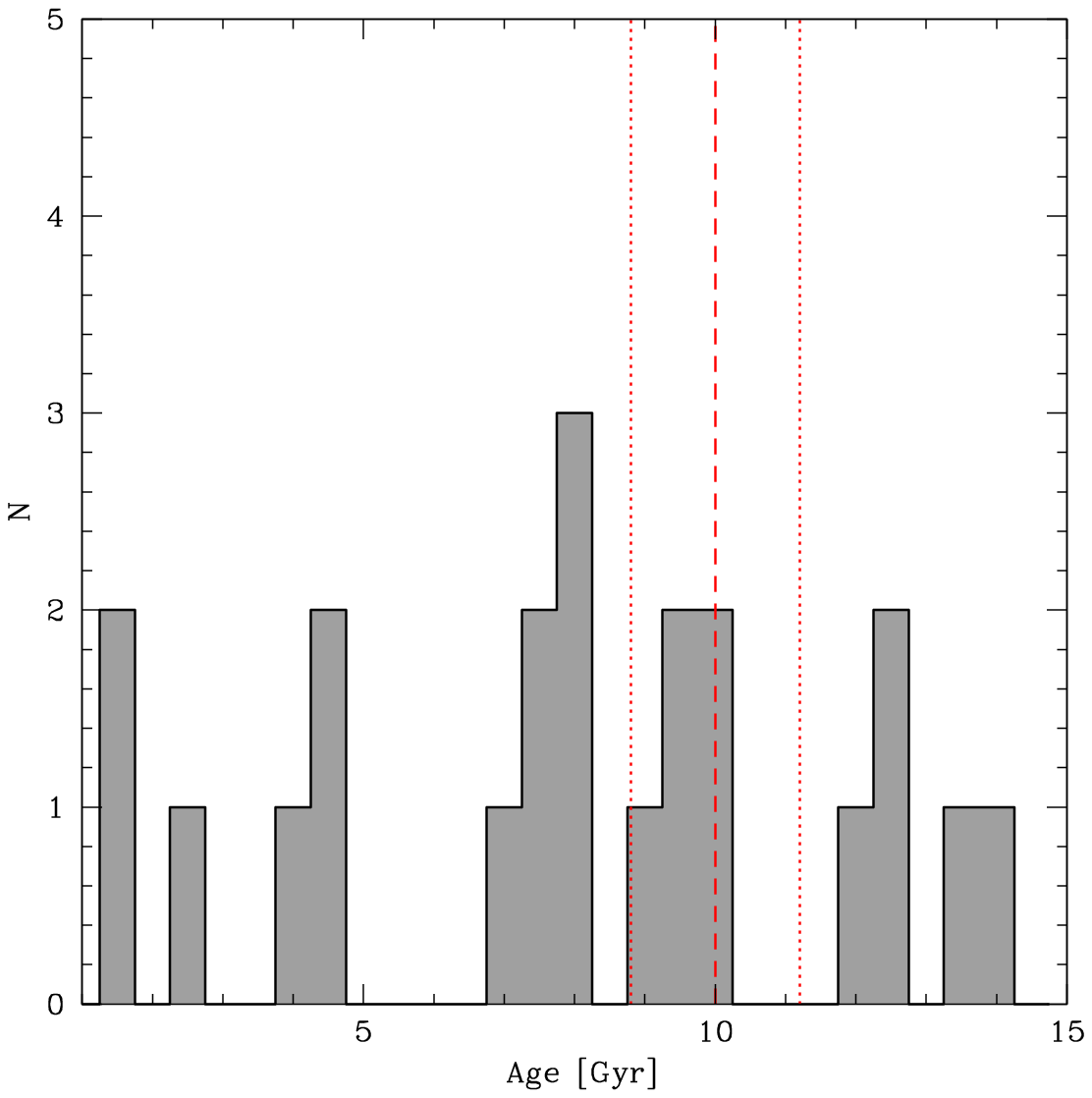}
    \includegraphics[width=9cm]{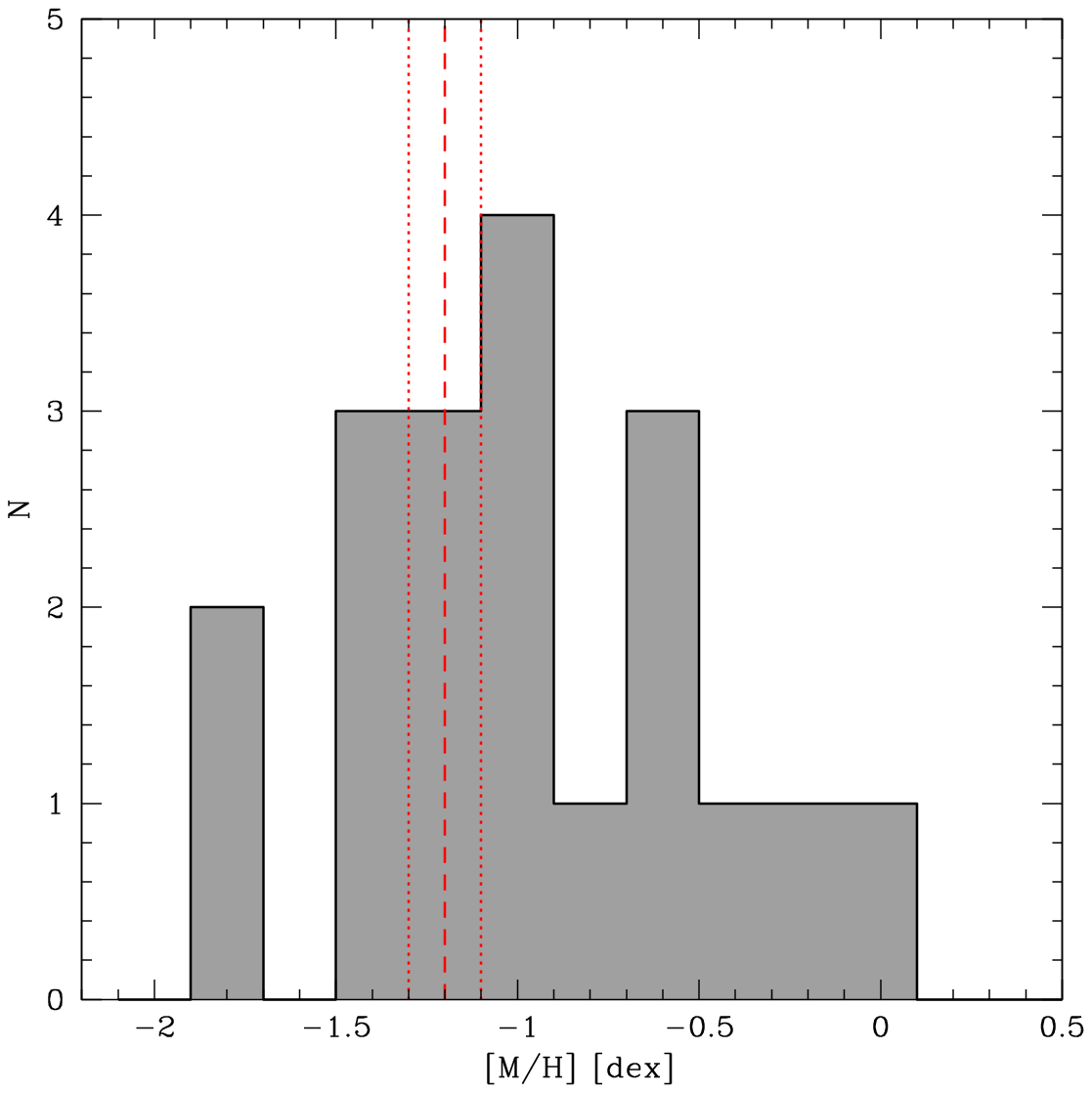}
    \caption{Stellar kinematics and stellar populations for UDGs from the literature compared with UDG11. Histograms (in grey) represent the stellar mass (top-left), velocity dispersion (top-right), age (bottom-left) and metallicity (bottom-right) for UDGs from previous works 
    \citep{Ruiz-Lara2018,Ferre-Mateu2018,Chilingarian2019,vanDokkum2019,Emsellem2019,Fensch2019,Gannon2021,Gannon2023,Toloba2018,Martin-Navarro2019}. 
    In all panels, the average value derived for UDG11 and error estimates are marked with the vertical red dashed and dotted lines, respectively. Values are obtained from the fit of the stacked spectrum inside $1R_{\rm e}$, and listed in the first column of Table~\ref{tab:UDG11}.} 
    \label{fig:conf_hist}
\end{figure*}

In Fig.~\ref{fig:conf_hist} we show the histograms of the stellar mass, velocity dispersion, 
age and metallicity measured for UDGs in previous works. 
The observed values obtained for UDG11 in this paper are also reported for comparison. 
These plots suggest that the velocity dispersion estimated for UDG11 ($\sigma \sim 20$~km/s)
fits into the broad range of $\sigma$ values for the majority of UDGs ($8\leq \sigma \leq32$~km/s), 
which peaks at $\sigma \sim 20$~km/s (see top-right panel of Fig.~\ref{fig:conf_hist}).
Based on the few measurements for the stellar population in UDGs, both age and metallicity span a wide range of values. 
On average, UDGs seem to be old ($7 \leq$~Age~$\leq 14$~Gyr) and metal-poor ($-1.8 \leq [M$/H$] \leq -0.5$~dex).
However, younger ages ($\sim 1$-4~Gyr) and sub-solar metallicity ([$M$/H]$ \sim -0.1$~dex) 
are found for a few objects (see lower panels of Fig.~\ref{fig:conf_hist}). 
The age and metallicity estimated for UDG11 fit well with 
the typical values observed for old and metal-poor UDGs. 

\begin{figure*}
    \includegraphics[width=8.95cm]{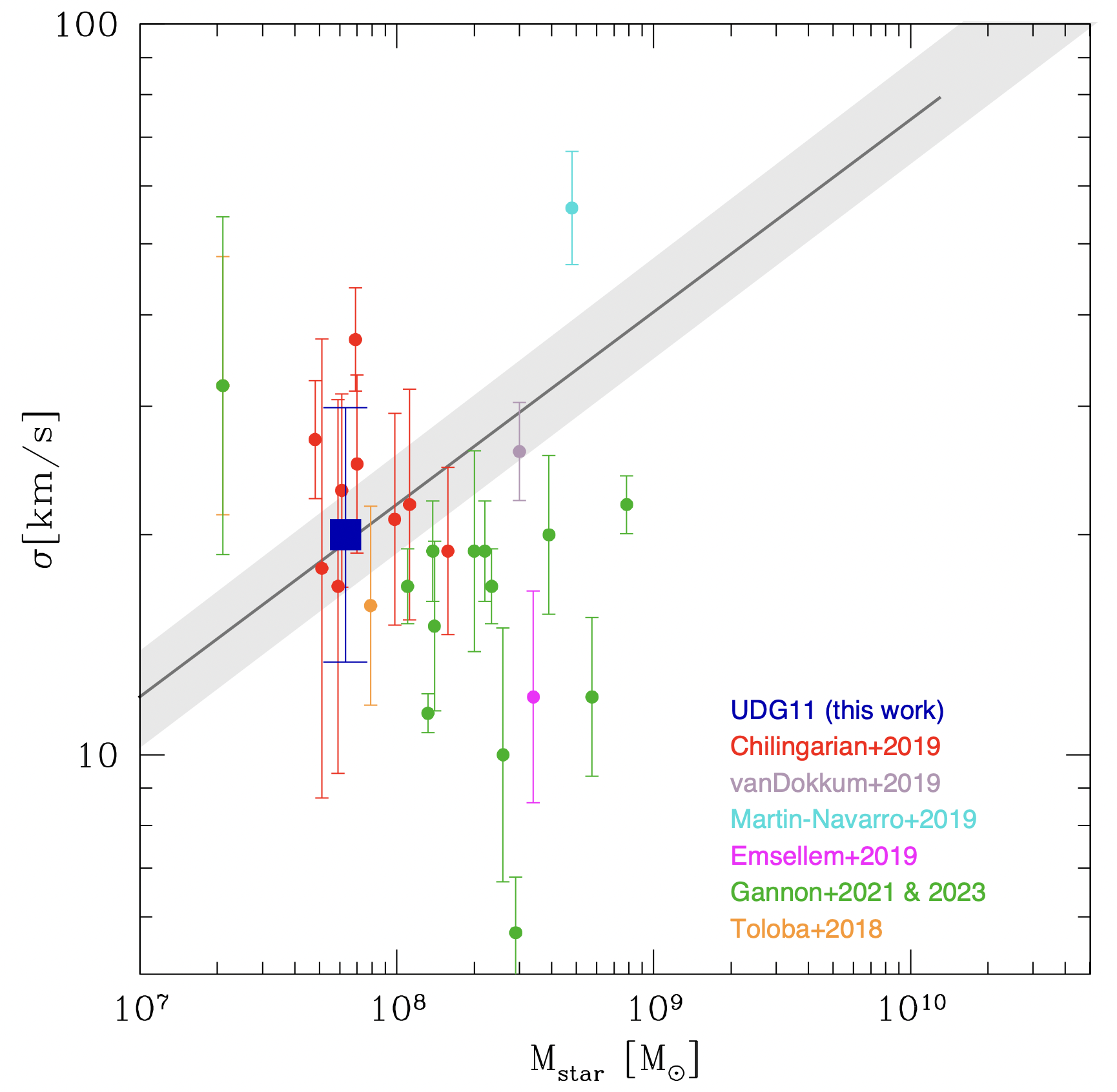}
    \includegraphics[width=9.3cm]{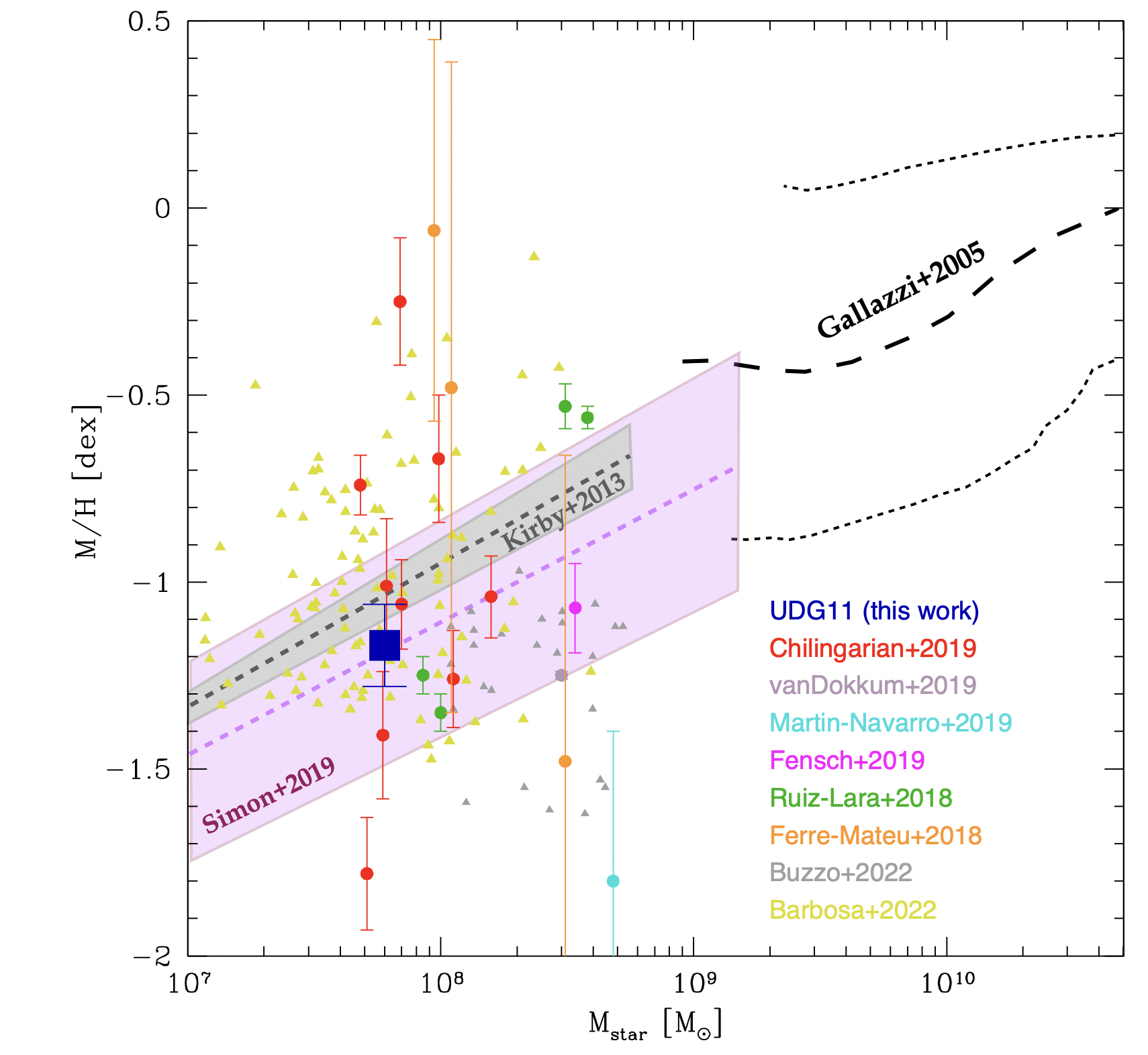}
    \caption{Velocity dispersion and metallicity as function of stellar mass for UDGs.
    Left panel - $\sigma$-mass relation for UDGs from the literature (coloured points) and UDG11 (blue square). 
     The grey area corresponds to the observed baryonic Tully-Fisher relation \citep{McGaugh2012}, obtained by scaling the rotation velocity by $\sqrt{3}$.
     Right panel - {   Mass-metallicity relation from observations, including
    UDG11. The mass-metallicity relations found for dwarf galaxies by \citet{Kirby2013} and by \citet{Simon2019} are marked in grey and purple regions, respectively. The average mass-metallicity relation for bright galaxies is marked by the black dashed line, and its range by the two dotted lines \citep{Gallazzi2005}. The metallicity derived by fitting the spectral energy distribution by \citet{Buzzo2022} and \citet{Barbosa2020a} are marked with grey and yellow triangles, respectively.}
    In both panels, data-points from \citet{Chilingarian2019} are UDG-like, since only two objects of the sample have $\mu_0 \geq 24$~mag/arcsec$^2$ and $R_{\rm e} \geq 1.5$~kpc.}
    \label{fig:sigZ_mass}
\end{figure*}

In Fig.~\ref{fig:sigZ_mass} we show the relations between the velocity dispersion and metallicity 
as a function of the stellar mass for UDGs.
In both panels, UDG11 has comparable values with other UDGs of similar stellar mass, taking into account 
the uncertainties on both measurements.
As already pointed out by \citet{Gannon2021}, in the $\sigma$-mass relation, the majority of UDGs, 
including UDG11, are found in the same region where other dwarf galaxies are located. 
However, some UDGs scatter to lower and higher $\sigma$ values  compared to the tighter $\sigma$-mass relation of ``normal'' dwarf galaxies.

In the mass-metallicity plane, UDGs show a large scatter (see right panel of Fig.~\ref{fig:sigZ_mass}). 
Most UDGs, including UDG11, are consistent with the typical values for dwarf galaxies in the same range of 
masses \citep{Kirby2013,Simon2019}. 
Also in this case, lower ([$M$/H]$\leq -1.5$) and higher  ([$M$/H]$\geq -1$) metallicity 
than the average are found. 
In particular, these estimates are also consistent with the observed age and metallicity derived 
for a large sample of dwarf galaxies \citep{Heesters2023,Romero-Gomez2023}, where
ages and metallicities range from 5 to 14 Gyr, and 0 to -1.9 dex, respectively.

A detailed discussion on the formation history of UDG11 is out of the scope of 
this paper, which, instead, will be presented in the context of the full sample.
However, we briefly comment below on how the LEWIS data can be used to address the structure and formation
of UDGs in a cluster environment. 
According to \citet{Sales2020}, at a given stellar mass, the T-UDGs show lower velocity dispersion 
and higher metallicity than the B-UDGs and normal dwarf galaxies. In addition, B-UDGs are found at larger 
cluster-centric distances ($\geq 0.5 R_{200}$) than T-UDGs, whose spatial distribution peaks around the 
cluster core \citep[see Fig.7 in][]{Sales2020}.
The observed properties of UDG11, from the deep VST images and LEWIS data, are similar to  
those predicted for the B-UDGs. UDG11 is located on the west side of the cluster, at about $0.4 
R_{\rm vir}$, in the low-density region of dwarf galaxies \citep[see][]{LaMarca2022b}, and the measured
velocity dispersion and metallicity are similar to those of dwarf galaxies (see Fig.~\ref{fig:sigZ_mass}).

\subsection{Dark matter content of UDG11}\label{sec:DM}

We computed the dynamical mass of UDG11 using the mass estimator proposed by \citet{Wolf2010},
where $M_{1/2} = 4 \times R_{\rm e} G^{-1} \langle\sigma\rangle_{\rm e}^2$. 
Using $\langle\sigma\rangle_{\rm e}=20\pm8$~km/s, derived from the fit of the 1D stacked spectrum,
we obtain $M_{1/2} =4.2 \pm0.8 \times 10^8$~M$_{\odot}$. 
This is the dynamical mass inside half of the total luminosity.
The absolute magnitude in the $V$ band for UDG11 is $M_{V,0}=-14.6\pm0.7$~mag, therefore
the total dynamical mass-to-light ratio is $M/L_V = 2\times M_{1/2}/L_V\simeq 14$~M$_{\odot}/$L$_{\odot}$.

UDG11 seems to have a total mass comparable to Local Group dwarfs of similar luminosity, 
which have a 
$M/L_V \sim 10 - 100$ \citep{Battaglia2022}. This suggests a dwarf-like DM halo for this UDG.

According to the stellar mass-halo mass relation 
derived down to the lowest stellar masses \citep{Wang2021}, which is comparable to  
the stellar mass of UDG11 ($M_{*}\sim 10^8$~M$_{\odot}$), the expected halo mass is M$_{\rm h}\sim 10^{10}$~M$_{\odot}$. {  A similar value is also obtained by using the halo mass-stellar mass relation
derived by \citet{Zaritsky2023}, where $M_{\rm h}=10^{10.35} \times [M_{*}/(10^8$~M$_{\odot})]^{0.63} \simeq 1.7 \times 10^{10}$~M$_{\odot}$.}

Using the scaling relation between the halo mass and the total number of GCs, which is
$\log[M_{\rm h}] = 9.68+1.01\times \log[N_{\rm GC}]$ \citep{Burkert2020}, 
and assuming that it is still valid in the low-mass regime, a consistent value of halo mass is found,
$M_{\rm h} \sim 4\times 10^{10}$~M$_{\odot}$, where we adopt $N_{\rm GC} = 5.9^{+2.2}_{-1.8}$ (see Sec.~\ref{sec:GCs}). 
{  However, it is worth noting here that this estimate for the halo mass is made under
the assumption that the central GC1 is a nuclear star cluster, and that GC2 does not belong to UDG11. 
Including both GCs would result in a higher $S_N$ and halo mass, which would still be consistent
within the uncertainties with the values quoted in Sec.~\ref{sec:GCs}.}

\section{Summary and concluding remarks}\label{sec:concl}

In this paper, we have presented the LEWIS project. 
LEWIS is a large observing programme started in 2021 to obtain the first homogeneous integral-field 
spectroscopic survey of LSB galaxies, including UDGs, in the Hydra\,I cluster of galaxies, with MUSE at ESO-VLT.
The LEWIS sample consists of 30 galaxies, 22 UDGs and 8 LSB dwarfs, with effective surface brightnesses in the range 
$25 \leq \mu_{\rm e} \leq 27.5$~mag/arcsec$^2$ in the $g$ band (Table~\ref{tab:UDGsample}). 
The total integration time per target varies from 2 hours, for the brightest objects, up to 6 hours for the 
faintest ones.

With the LEWIS project we expect to constrain
 {\it i)} the fraction of baryonic versus DM content, {\it ii)} the star formation history, 
 and {\it iii)} the GCs content by means of spectroscopic $S_N$, in all sample galaxies.
Given the large variety of observed properties for UDGs (mainly based on deep images),
and theoretical scenarios, which envisage various possibilities for the formation of this kind of galaxy, 
the main outcomes of this project are a notable boost to our knowledge of the UDG structure and formation. 


LEWIS observations are still ongoing. In this paper we have presented the LEWIS sample 
and the measured redshift for the {  20 targets (16 UDGs and 4 LSB dwarfs) observed so far. 
Fourteen of them have been confirmed as Hydra~I cluster members, with systemic velocities ranging from 
3483~km/s to 4302~km/s, 
which correspond to $\sim$1$\sigma$ from the cluster mean velocity ($V_{\rm sys} = 3683 \pm 46$~km/s, $\sigma_{\rm cluster} \sim 700$~km/s). Three galaxies of the sample have systemic velocities within $2 \sigma_{\rm cluster}$, therefore 
they might be still considered as cluster members.
Finally, the remaining three galaxies with larger or smaller systemic velocities could be considered as
background or foreground galaxies, respectively.}

To assess the quality of the LEWIS data, we have also described the data analysis that we adopt in this project, 
for one of the sample galaxies, UDG11, chosen as test case. 
UDG11 is located far from the cluster core,
on the west side, at a distance of $\sim0.4$~R$_{\rm vir}$. It has an absolute
$r-$band magnitude of $M_r = -14.75$~mag and a stellar mass of $0.63\times 10^8$~M$_\odot$.
The MUSE data obtained for this target have a total integration time of 6.10 hours, which allowed us to
obtain a SNR$\sim16$ per spaxel, for the 1D stacked spectrum inside $1 R_{\rm e}$.
For this target, we have derived the stellar kinematics, constrained the average 
age and metallicity of the stellar populations, estimated the total number of spectroscopically confirmed GCs,
and provided the dynamical mass. All the above quantities are compared with the available measurements for
UDGs and dwarf galaxies in literature. Results are summarised below.

\begin{itemize}
    \item By fitting the stacked spectrum inside $1R_{\rm e}$, on the full MUSE rest-frame wavelength range 
($4800 - 9000$ \AA), we obtained a velocity dispersion $\sigma =20 \pm8$~km/s, 
 a metallicity [$M$/H]$=-1.17 \pm 0.11$~dex and an age of $10 \pm 1$~Gyr.

\item The spatially resolved stellar kinematics, obtained from the Voronoi-binned spectra with SNR = 10, show that UDG11 does not show a significant gradient of $V_{\rm sys}$ and $\sigma_{\rm LOS}$ along major 
and minor photometric axes. The mean value of the velocity dispersion is $\langle\sigma\rangle_{\rm e}=27 \pm 8$~km/s, which is consistent with the estimate from the fit of the stacked spectrum inside $1R_{\rm e}$.

\item The absence of emission lines in the MUSE cube for UDG11 
suggests that this galaxy lacks ionised gas.

\item Four point sources in the UDG11 cube have radial velocities consistent with the Hydra\,I cluster.
Two of those four sources appear kinematically associated to UDG11 (with relative velocities $\sim$10-30~km/s).
The other two sources, 
with larger velocities, are classified as intra-cluster GCs. 
When corrected for our spectroscopic measurement magnitude limit, we obtain an estimated 
total number of GCs $N_{\rm GC}=5.9^{+2.2}_{-1.8}$, and the corresponding specific frequency is 
$S_{\rm N}= 8.4^{+3.2}_{-2.7}$. This is consistent with the typical $S_N$ estimates 
for dwarf galaxies of similar luminosity \citep{Lim2018}.

\item The stellar velocity dispersion, age and metallicity derived for UDG11 are comparable with values
derived for other UDGs of similar stellar masses, estimated in previous works 
(see Fig.~\ref{fig:conf_hist}).

\item {  The total mass inside $1R_{\rm e}$ and the total dynamical mass-to-light ratio estimated for UDG11
are $M_{1/2} \simeq 4\times 10^8$~M$_{\odot}$ and $M/L_V \simeq 14 M_{\odot}/L_{\odot}$, respectively}. 
These estimates are 
comparable to that of Local Group dwarfs of similar total luminosity \citep{Battaglia2022}, 
suggesting a dwarf-like DM halo for this UDG.

\end{itemize}

In summary, we found that UDG11 is old, metal-poor and has a DM content 
comparable to those observed for dwarf galaxies. Being also gas-poor, all these observed properties might suggest that an external process acted in the past to form UDG11, 
and that the formation channels based on internal mechanisms
could be reasonably excluded (see Sec.~\ref{sec:intro} and Fig.~\ref{fig:formation}).
In particular, UDG11 might have formed as a LSB which then lost its gas and was quenched 
when it was accreted by the Hydra\,I cluster. 

These results have further proven the power of IF spectroscopic data, and in particular of the 
MUSE at ESO-VLT, to study the structure of LSB galaxies and
to constrain the formation history of UDGs.
To date, similar studies are available only for about 35 UDGs in total, 
mainly in the Coma cluster, because of the challenging observations at such very LSB levels. 
In particular, IF spectroscopy is available 
only for a dozen UDGs \citep{Emsellem2019,Muller2020,Gannon2021,Gannon2023}. 
With LEWIS we will make a decisive impact in this field. 
We will double the number of spectroscopically 
studied UDGs and establish, for the first time, properties of a (nearly) complete sample of UDGs 
in a galaxy cluster, Hydra I. This cluster is currently undergoing an active assembly phase, 
thus offering a range of environments for its members. 
\begin{acknowledgements}
Based on observations collected at the European Southern Observatory under ESO programmes 108.222P.001, 108.222P.002, 108.222P.003.
{  We thank the anonymous referee for their useful suggestions that helped to improve the paper.}
Authors wish to thank Dr. Lodovico Coccato (ESO-Garching) 
for the kind support during the MUSE data reduction.
{  E.I. wish to thank L. Buzzo, P.-A. Duc, A. Ferre-Mateu, 
J. Gannon, F. Marleau, O. Muller and R. Peletier for the useful comments and discussions on the work presented in this paper.}
E.I. acknowledges support by the INAF GO funding grant 2022-2023.
JH wishes to acknowledge CSC – IT Center for Science, Finland, for computational resources.
EMC is supported by MIUR grant PRIN 2017 20173ML3WW-001 and Padua University grants DOR2019-2021.
J.~F-B  acknowledges support through the RAVET project by the grant PID2019-107427GB-C32 from the Spanish Ministry of Science, Innovation and Universities (MCIU), and through the IAC project TRACES which is partially supported through the state budget and the regional budget of the Consejer\'ia de Econom\'ia, Industria, Comercio y Conocimiento of the Canary Islands Autonomous Community.
MAR acknowledges funding from the European Union’s Horizon 2020 research and innovation programme under the Marie Skłodowska-Curie grant agreement No 101066353 (project ELATE).
KF acknowledges support through the ESA Research Fellowship programme.
Authors acknowledge the use of the {\it dfitspy} tool \citep{Romain2019} and MUSE Python Data Analysis Framework (MPDAF) \citep{Bacon2016}.
\end{acknowledgements}

\bibliographystyle{aa.bst}
\bibliography{LEWIS.bib}


\begin{appendix}\label{sec:appendix}

\section{MUSE reconstructed images and stacked spectra}\label{sec:images}
In this section, for each target observed in the LEWIS sample (see Table\ref{tab:UDGsample}), 
we show the reconstructed images from the MUSE cube and the stacked spectrum inside $1R_{\rm e}$, 
including the best fit to derive the redshift estimate.
All of them are obtained using the standard data-reduction pipeline with some optimized parameters, as described in Sec.~\ref{sec:skysub}.


\section{Error estimates: general overview}\label{sec:errors}

The uncertainties present in the analysis are related to the following sources of errors:

\begin{enumerate}
\item Quality of the data (noise and residual sky lines)
\item Uncertainty on the spectral resolution
\item Prior on the regularization
\item Choice of the spectral mask
\item Calibration errors and selection of multiplicative and additive polynomials
\end{enumerate}

\noindent Here we describe how we estimate the level of error from these sources of uncertainties:

\begin{enumerate}
\item Our MUSE spectrum for UDG11 has a per pixel SNR of 16. This significant amount of noise comes 
from two sources. The statistics of counting photons, which can be described with a Poissonian 
function, and a systematic source that includes uncorrected calibration problems, sky line residuals, 
and other imperfections of the reduction pipeline. While the Poissonian noise is not usually 
a source of bias, systematic noise can often present severe bias effects if not properly addressed, 
as its effect on the spectrum can be easily confused as part of the signal.

We have treated most of the systematic effects using an improved datacube reduction pipeline (see Sec.~\ref{sec:skysub}) and proper masking of noisy and contaminated regions in the analysed spectra (see e.g., Fig.~\ref{fig:UDG11_1Respectra}), improving the effective SNR to $\sim$20 per pixel. Therefore, we only benefit from an upper limit on possible biases. We give such an estimate using simulations in Sec.~\ref{sec:data_quality}, and unless stated, we henceforth assume that most of the errors are dominated by Poisson statistics.

\item An estimate of the typical MUSE's LSF has been already measured by \cite{Bacon2017} and \cite{Emsellem2019}. 
We have used the equation given by \citet{Bacon2017} to approximate the MUSE's LSF under 
the assumption that at the best FWHM achieved by the data, the templates never significantly ($\Delta$FWHM $>0.1$ \AA) fall below 
the resolution of the MUSE spectra. From \cite{Bacon2017}, we expect a maximum deviation from the true FWHM at any point 
of the wavelength axis of about $\Delta$FWHM $=0.1$ \AA. This would translate to an uncertainty of $<10$ km/s as shown in Fig.~\ref{fig:test_sigma}.

\item Regularization (stellar populations only): we used the suggested pPXF method to select a regularization parameter depending on the data noise, which is tied to the SNR. In this sense, regularization is not an independent source of uncertainty.

\item Choice of spectral mask: Minor changes on the spectral mask selection can cause small variations on the resulting parameters, therefore, we change it randomly for our error estimation described in Appendix~C.

\item Calibration errors: Usually corrected by using multiplicative and additive polynomials. A study of the effect of the additive polynomial degree for this particular source has been already presented in the main text of this study (see Fig.~\ref{fig:test_poly_sigma}). Regarding the effect of Legendre multiplicative polynomials, we have ran several tests to address this effect. We verify that at degrees $8-11$ there are virtually no differences in the total metallicity and median age of the recovered stellar population, but the differences grow up to $0.1$ dex in [$M$/H] and $1$ Gyr in stellar age when using a very low ($5-7$) or high ($>12$) degree. We therefore place a strong prior on a Legendre multiplicative polynomial degree of 10.

To summarize, we have found that when properly constrained, the above effects have a similar contribution to the total uncertainty.

\end{enumerate}

\section{Error bar estimation}\label{sec:error_bars}

\subsection{On the velocity dispersion}
To determine the error bars on the systemic velocity and velocity dispersion of the $1R_{\rm e}$ stacked spectrum, we decided to apply a bootstrap-like procedure: First, we generate different variations of said stacked spectrum by removing and replacing random regions (with spaxels surrounding $1R_{\rm e}$) within the extraction zone. This corresponds to $10\%$ of all stacked spaxels. Second, we re-fit the spectrum introducing Poissonian noise without drastically reducing the SNR. Effectively, we introduce Poissonian noise that introduces a standard deviation on the SNR per pixel distribution of $<3$. Finally, we fit the generated spectra via exploring a grid of randomly generated input parameters (described in Appendix~B), emulating common Monte Carlo approaches. In all cases, we do not perturb the polynomial degrees used as described in Sec.\ref{sec:sfh}. Our results show that the velocity dispersion measurement on the stacked spectrum has an uncertainty of 10 km/s, which agrees with our simulations (Appendix~\ref{subsec:simulations}).

\subsection{On the stellar population}
To get age and metallicity uncertainties, we apply the same methodology as above but with the inclusion of mild perturbations on the regularization parameter as a function of S/N, in addition to fixing the velocity dispersion values (within 1$\sigma$) to those obtained in the previous step. 

\section{Simulations}\label{subsec:simulations}

\subsection{Signal-to-noise limitations}\label{sec:poisson_tests}
As described in Sec.~\ref{sec:kin}, from the best fit of the stacked spectrum inside $1 R_{\rm e}$ of UDG11
we obtained a velocity dispersion of $\sim 20$~km/s, which is lower than the spectra resolution of 
MUSE. In addition, we aim at measuring  metallicities with uncertainties below $\sim$ 0.3. 
To achieve this, we require a clear understanding of the performance of pPXF and our pipeline in measuring 
small variations on the width of the main absorption lines with limited SNR. 
Therefore, based on the E-MILES models, we simulate mock UDG11-like spectra to study how the 
retrieved quantities from the pPXF fit vary with the SNR of the spectrum. 
Firstly, we take a stellar population with log(Age)$=$10 and [$M$/H]$=-$1.2, and construct a synthetic spectrum based on it. Next, we convolve the spectrum with an approximated version of MUSE's line-spread function (LSF) as presented in Bacon et al. (2017). 
Finally, we further convolve it with varying kernels to simulate different velocity dispersions.
Subsequently, we introduced noise by generating several mock spectra with different signal-to-noise levels, ranging from 5 to 120 per pixel, by adding Poissonian noise. 
This allows us to evaluate the performance of the fit at very low SNR values. 
Fig.~\ref{fig:test_sigma} illustrate the results of this test.

The figures show an unbiased trend that performs relatively well at SNR $\sim 10-15$, 
when trying to retrieve velocity dispersion as low as $\sigma \sim$ 15 km/s. 
The error at this level is of about 10 km/s.

\subsection{Bias analysis}\label{subsec:bias}

In order to put a rough upper bound on the maximum degree of bias in our results, under the assumption that all of our noise is systematic, we extract a clean background region of the same size as the one used for the stacked 1D UDG11 spectrum, present on the same cube. Then, we repeat the procedures described in Sec.~\ref{subsec:simulations}, including the background spectrum (containing residual sky lines and other systematics) instead of the random noise. After 1000 simulations, we estimate a maximum bias of around $+0.1$ in metallicity, and a negligible one on the mean stellar age.

\begin{figure*}
    \includegraphics[width=18cm]{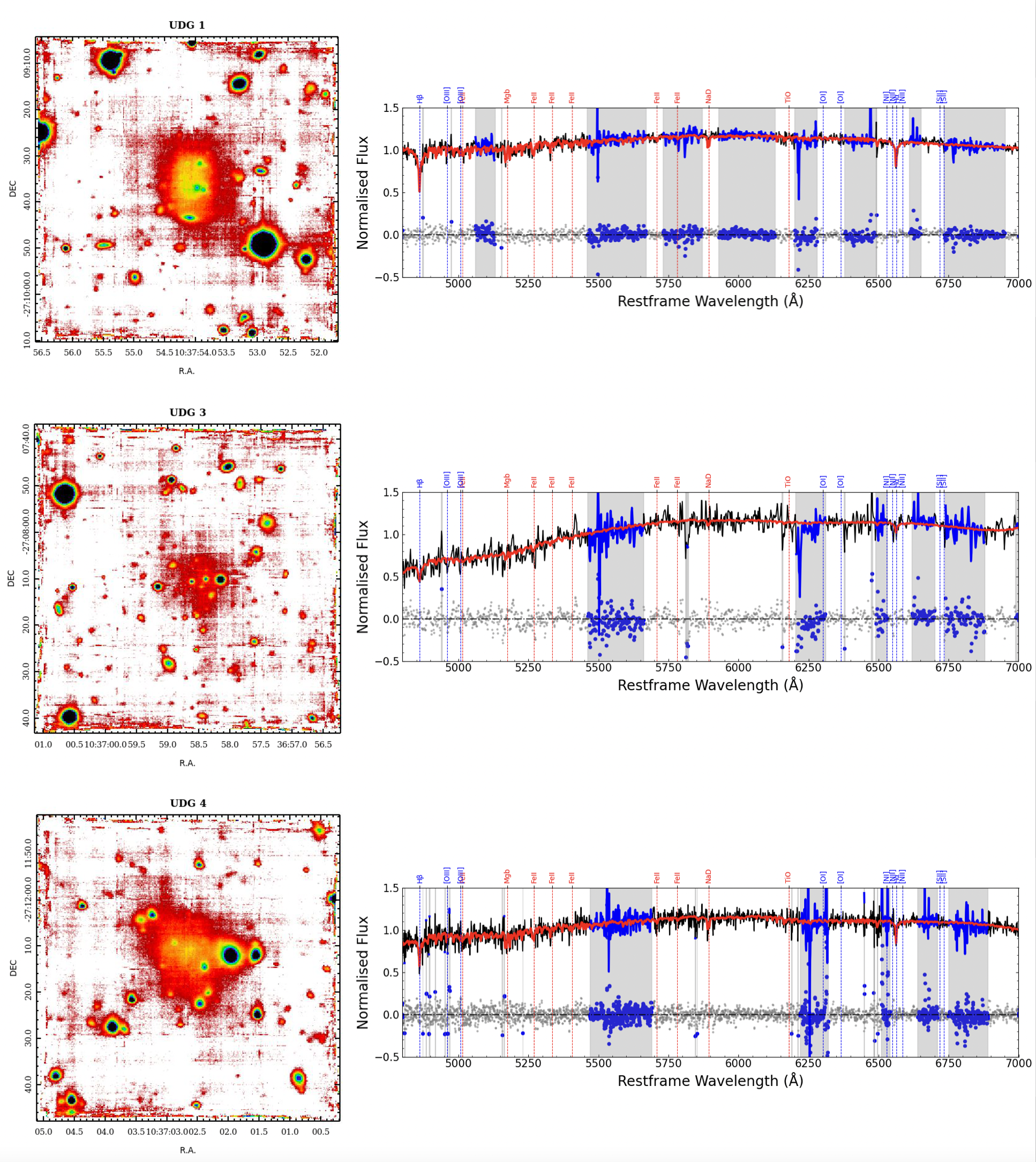}
    \caption{MUSE reconstructed images (left panels) and stacked spectra (right panels) of the observed LEWIS targets. From the top to the lower panels, UDG1, UDG3 and UDG4 are shown, respectively. The red line corresponds to the pPXF best fit using the set-up described in Sec.~\ref{sec:redshift}. Grey boxes mark the masked regions, excluded from the fit.}
    \label{fig:UDG_1_3_4}
\end{figure*}

\begin{figure*}
    \includegraphics[width=18cm]{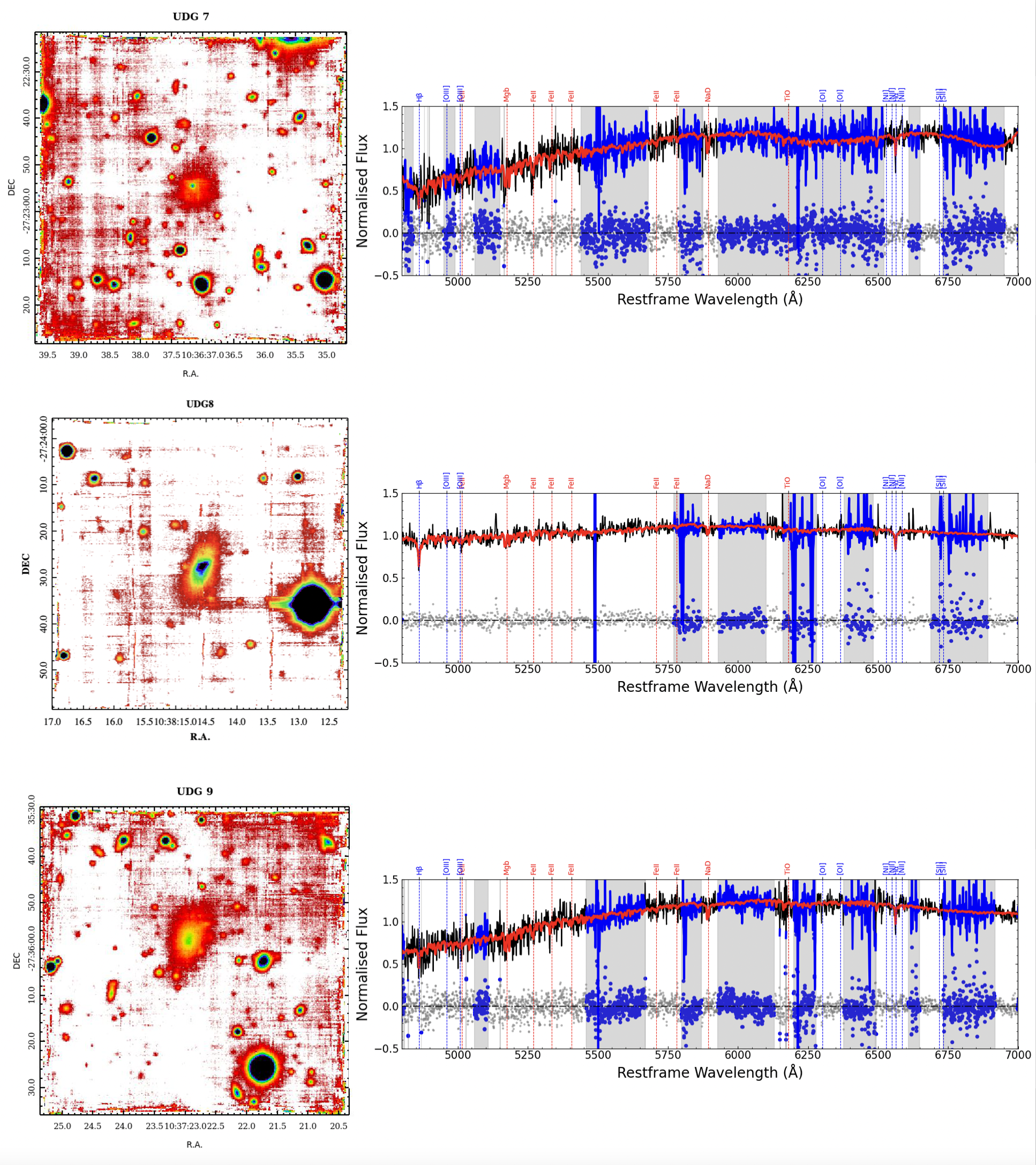}
    \caption{Same as Fig.~\ref{fig:UDG_1_3_4}, for UDG7, UDG8 and UDG9.}
    \label{fig:UDG_7_8_9}
\end{figure*}

\begin{figure*}
    \includegraphics[width=18cm]{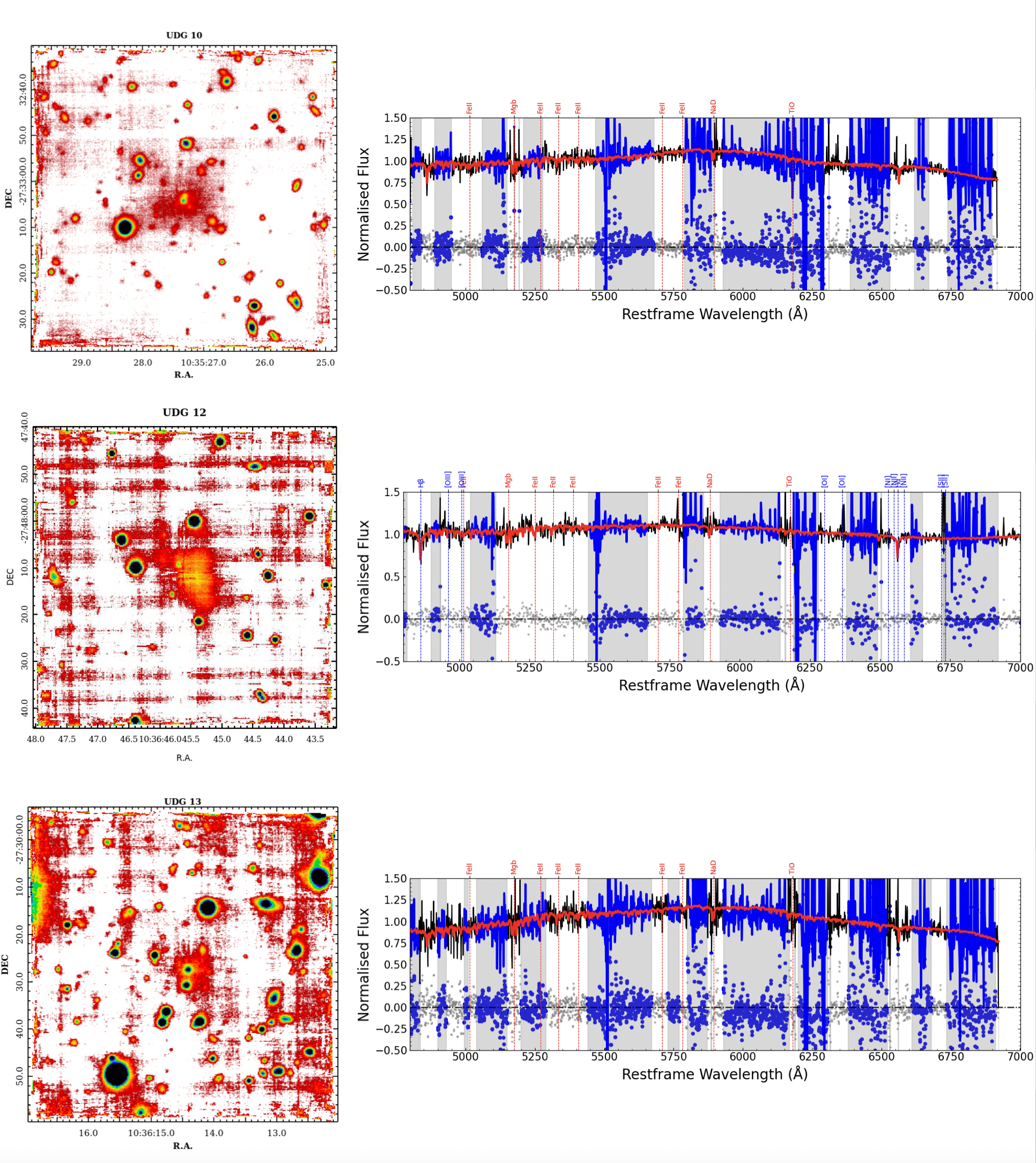}
    \caption{Same as Fig.~\ref{fig:UDG_1_3_4}, for UDG10, UDG12 and UDG13.}
    \label{fig:UDG_10_12_13}
\end{figure*}

\begin{figure*}
    \includegraphics[width=18cm]{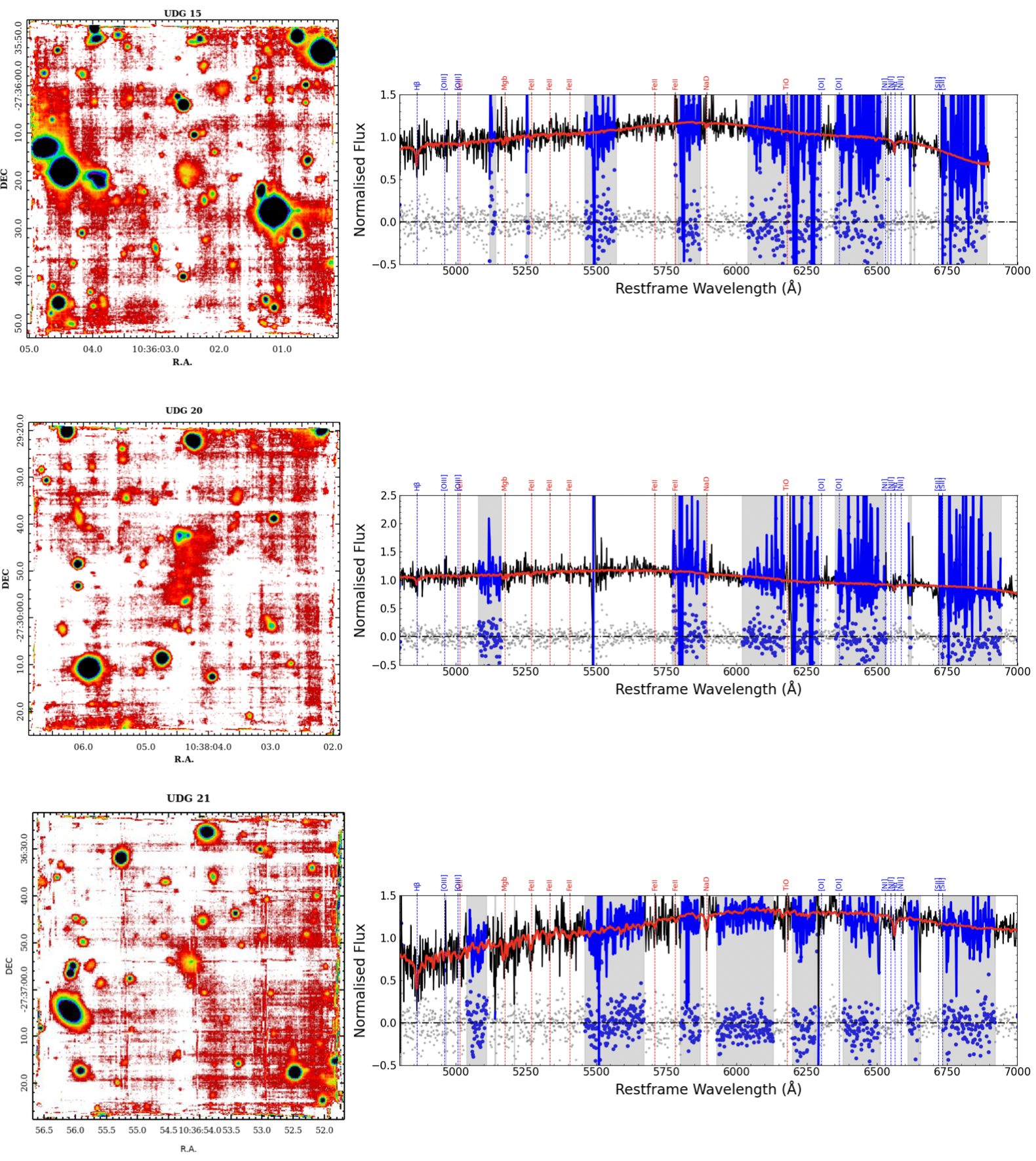}
    \caption{Same as Fig.~\ref{fig:UDG_1_3_4}, for UDG15, UDG20 and UDG21.}
    \label{fig:UDG_15_20_21}
\end{figure*}

\begin{figure*}
    \includegraphics[width=18cm]{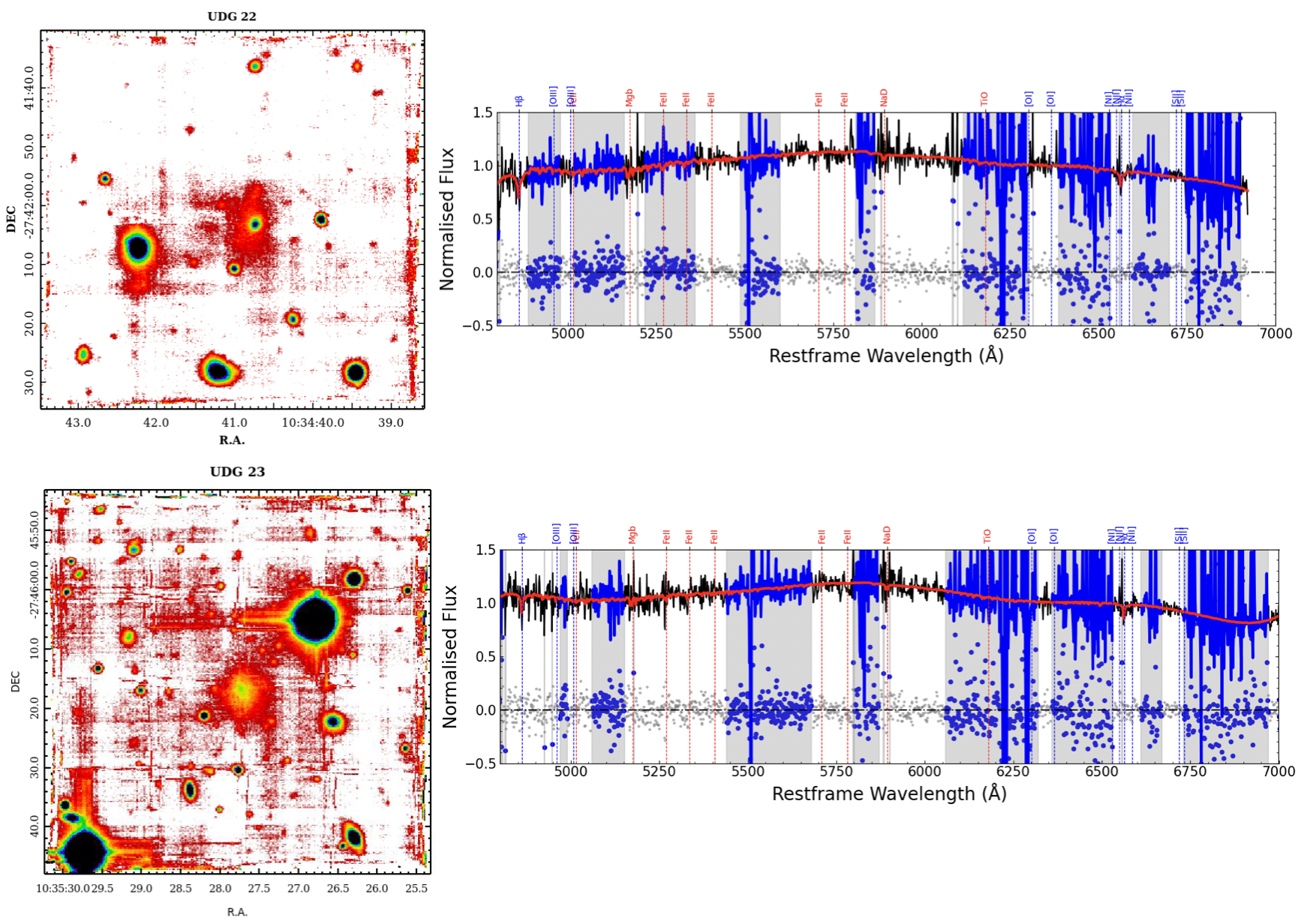}
    \caption{Same as Fig.~\ref{fig:UDG_1_3_4}, for UDG22 and UDG23.}
    \label{fig:UDG_22_23}
\end{figure*}

\begin{figure*}
    \includegraphics[width=18cm]{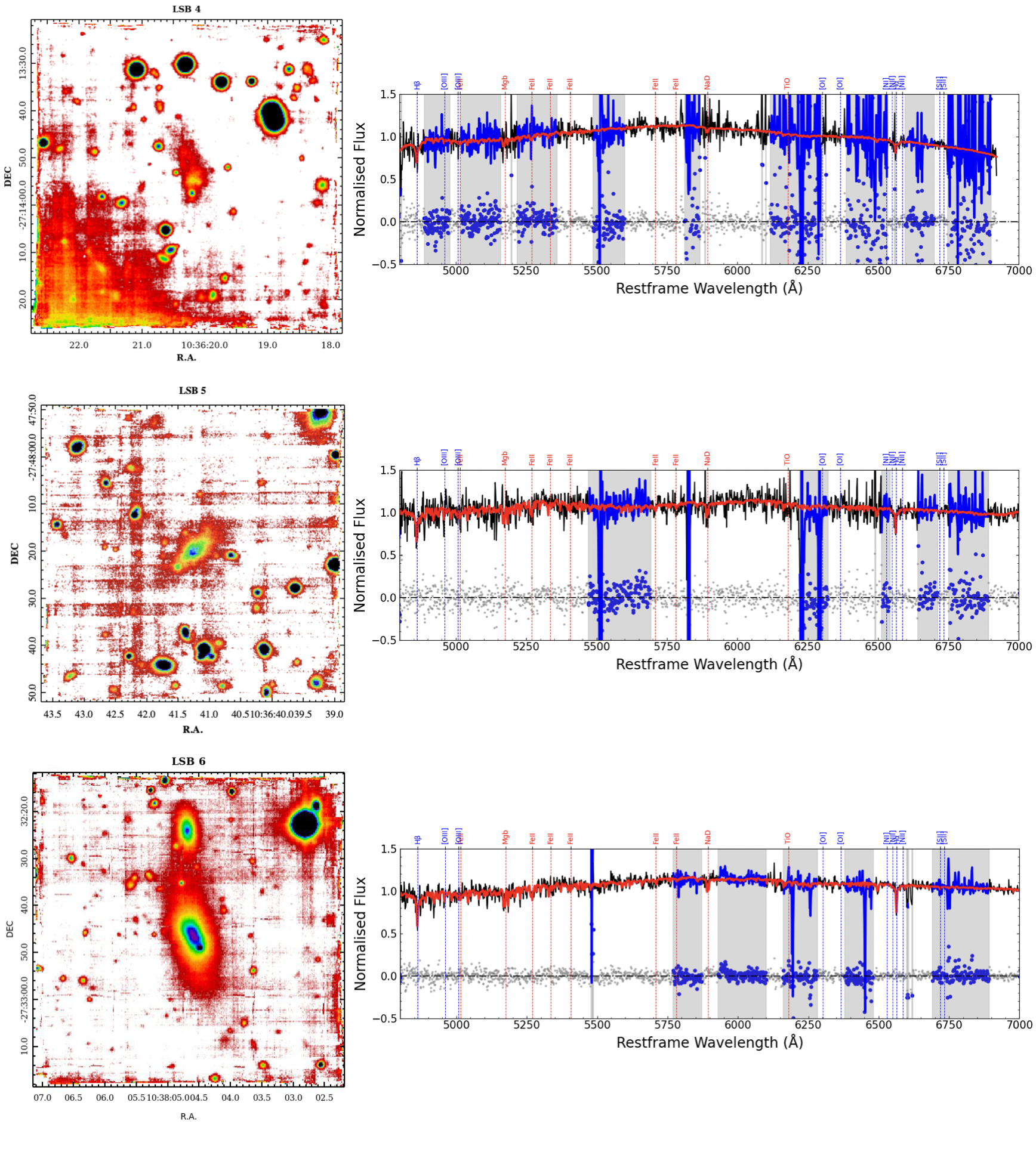}
    \caption{Same as Fig.~\ref{fig:UDG_1_3_4}, for LSB4, LSB5 and LSB6.}
    \label{fig:LSB_4_5_6}
\end{figure*}

\begin{figure*}
    \includegraphics[width=18cm]{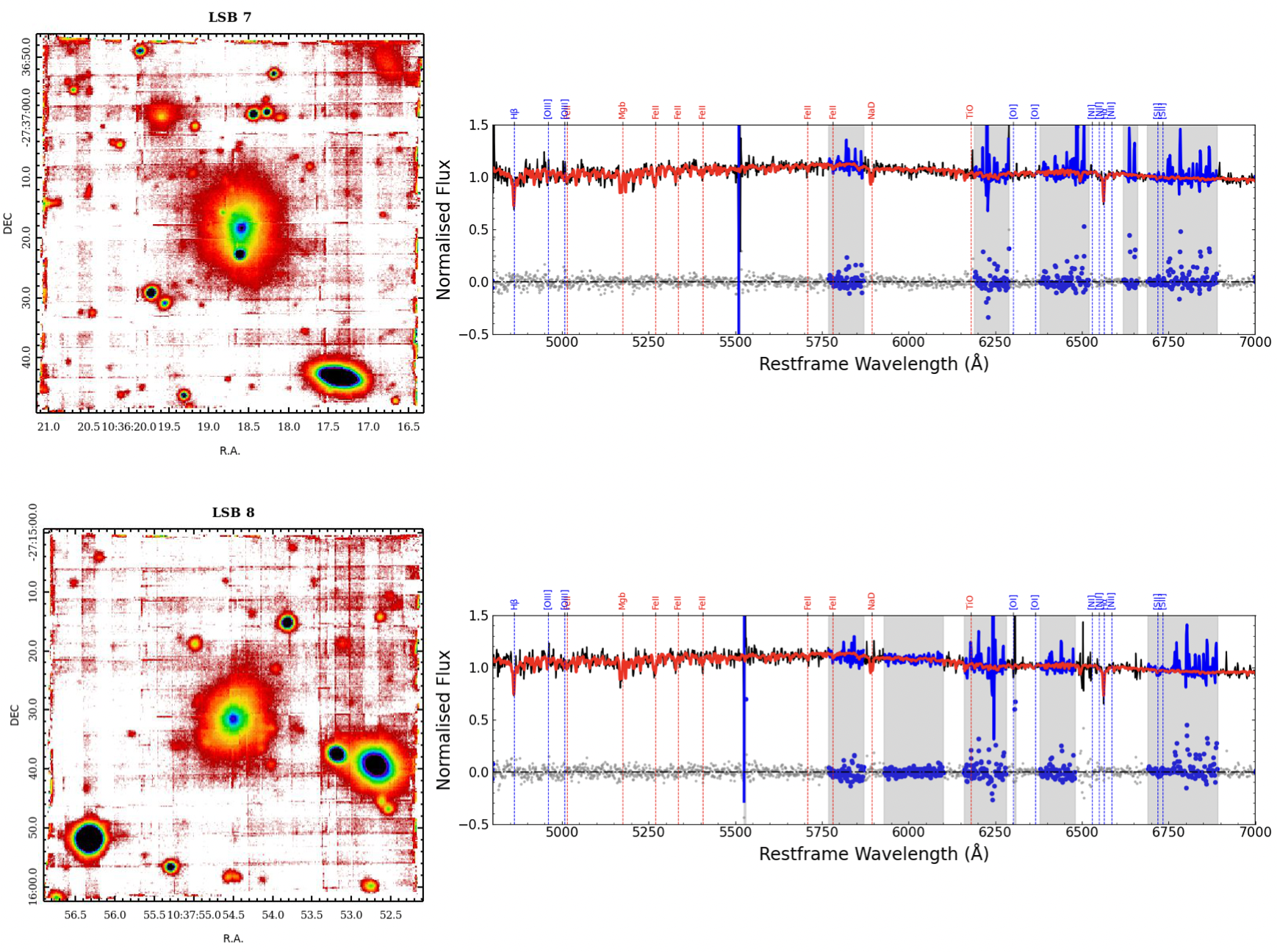}
    \caption{Same as Fig.~\ref{fig:UDG_1_3_4}, for LSB7 and LSB8.}
    \label{fig:LSB_7_8}
\end{figure*}

\end{appendix}
  
\end{document}